\documentclass[superscriptaddress,aps,amsmath,amssymb,floatfix,reprint,raggedbottom,longbibliography]{revtex4-1}
\usepackage{graphicx}
\usepackage[colorlinks=true,citecolor=blue,linkcolor=blue,urlcolor=blue]{hyperref}
\usepackage{amsmath}
\usepackage{dcolumn}
\usepackage{xcolor}
\usepackage{fancyhdr}
\usepackage{lipsum}
\usepackage{soul}

\usepackage{footnote}

\makeatletter 
\renewcommand{\fnum@figure}{\textbf{FIG.~\thefigure}}
\makeatother

\makeatletter
\def\bbordermatrix#1{\begingroup \m@th
  \@tempdima 4.75\p@
  \setbox\z@\vbox{%
    \def\cr{\crcr\noalign{\kern2\p@\global\let\cr\endline}}%
    \ialign{$##$\hfil\kern2\p@\kern\@tempdima&\thinspace\hfil$##$\hfil
      &&\quad\hfil$##$\hfil\crcr
      \omit\strut\hfil\crcr\noalign{\kern-\baselineskip}%
      #1\crcr\omit\strut\cr}}%
  \setbox\tw@\vbox{\unvcopy\z@\global\setbox\@ne\lastbox}%
  \setbox\tw@\hbox{\unhbox\@ne\unskip\global\setbox\@ne\lastbox}%
  \setbox\tw@\hbox{$\kern\wd\@ne\kern-\@tempdima\left[\kern-\wd\@ne
    \global\setbox\@ne\vbox{\box\@ne\kern2\p@}%
    \vcenter{\kern-\ht\@ne\unvbox\z@\kern-\baselineskip}\,\right]$}%
  \null\;\vbox{\kern\ht\@ne\box\tw@}\endgroup}
\makeatother

\begin{document}
\title{Stochastic p-bits for Invertible Logic}
\author{Kerem Yunus  Camsari}
\email{kcamsari@purdue.edu}      
\affiliation{School of Electrical and Computer Engineering, Purdue University, IN, 47907}
\author{Rafatul Faria}
\affiliation{School of Electrical and Computer Engineering, Purdue University, IN, 47907}
\author{Brian M. Sutton}
\affiliation{School of Electrical and Computer Engineering, Purdue University, IN, 47907}
\author{Supriyo Datta}
\email{datta@purdue.edu}      
\affiliation{School of Electrical and Computer Engineering, Purdue University, IN, 47907}
\date{\today}

\begin{abstract}
Conventional semiconductor-based logic and nanomagnet-based memory devices are built out of stable, 
deterministic units such as standard MOS (metal oxide semiconductor) transistors, or nanomagnets with energy barriers in excess of $\approx$ 40-60 kT. In this paper we show that unstable, stochastic units which we call ``p-bits'' can be interconnected to create robust correlations that implement \textit{precise Boolean functions} with impressive accuracy, comparable to standard digital circuits. At the same time they are \textit{invertible}, a unique property that is absent in standard digital circuits. When operated in the direct mode, the input is clamped, and the network provides the correct output. In the inverted mode, the output is clamped, and the network fluctuates among all possible inputs that are consistent with that output. First, we present a detailed implementation of an invertible gate to bring out the key role of a single three-terminal transistor-like building block to enable the construction of correlated p-bit networks. The results for this specific, CMOS-assisted nanomagnet-based hardware implementation agree well with those from a universal model for p-bits, showing that p-bits need not be magnet-based: any three-terminal tunable random bit generator should be suitable. We present a general algorithm for designing a Boltzmann machine (BM) with a symmetric connection matrix [J] ($J_{ij}=J_{ji}$), that implements a given truth table with p-bits. The [J] matrices are relatively sparse with a few unique weights for convenient hardware implementation. We then show how BM Full Adders can be interconnected in a \textit{partially directed} manner ($J_{ij} \ne J_{ji}$) to implement large logic operations such as 32-bit binary addition. Hundreds of stochastic p-bits get precisely correlated such that the correct answer out of $2^{33} \ (\approx 8 \ \rm billion)$ possibilities can be extracted by looking at the statistical mode or majority vote of a number of time samples. With perfect directivity ($J_{ji}$=0) a small number of samples is enough, while for less directed connections more samples are needed, but even in the former case logical invertibility is largely preserved. This combination of digital accuracy and logical invertibility is enabled by the hybrid design that uses bidirectional BM units to construct circuits with partially directed inter-unit connections.  We establish this key result with extensive examples including a 4-bit multiplier  which in inverted mode functions as a factorizer.

 \end{abstract}
\maketitle

\section{Introduction}
  \vspace{0em}

Conventional semiconductor-based logic and nanomagnet-based memory devices are built out of stable, deterministic units such as standard MOS (metal oxide semiconductor)  transistors, or nanomagnets with energy barriers in excess of $\approx$ 40-60 kT. The objective of this paper is to introduce the concept of what we call ``p-bits'' representing unstable, stochastic units which can be interconnected to create robust correlations that implement precise Boolean functions with impressive accuracy comparable to standard digital circuits. At the same time this ``probabilistic spin logic'' (PSL) is \textit{invertible}, a unique property that is absent in standard digital circuits. When operated in the direct mode, the input is clamped, and the network provides the correct output. In the inverted mode, the output is clamped, and the network fluctuates among all possible inputs that are consistent with that output.

\begin{figure*}[t!]
\includegraphics[width=.95\linewidth]{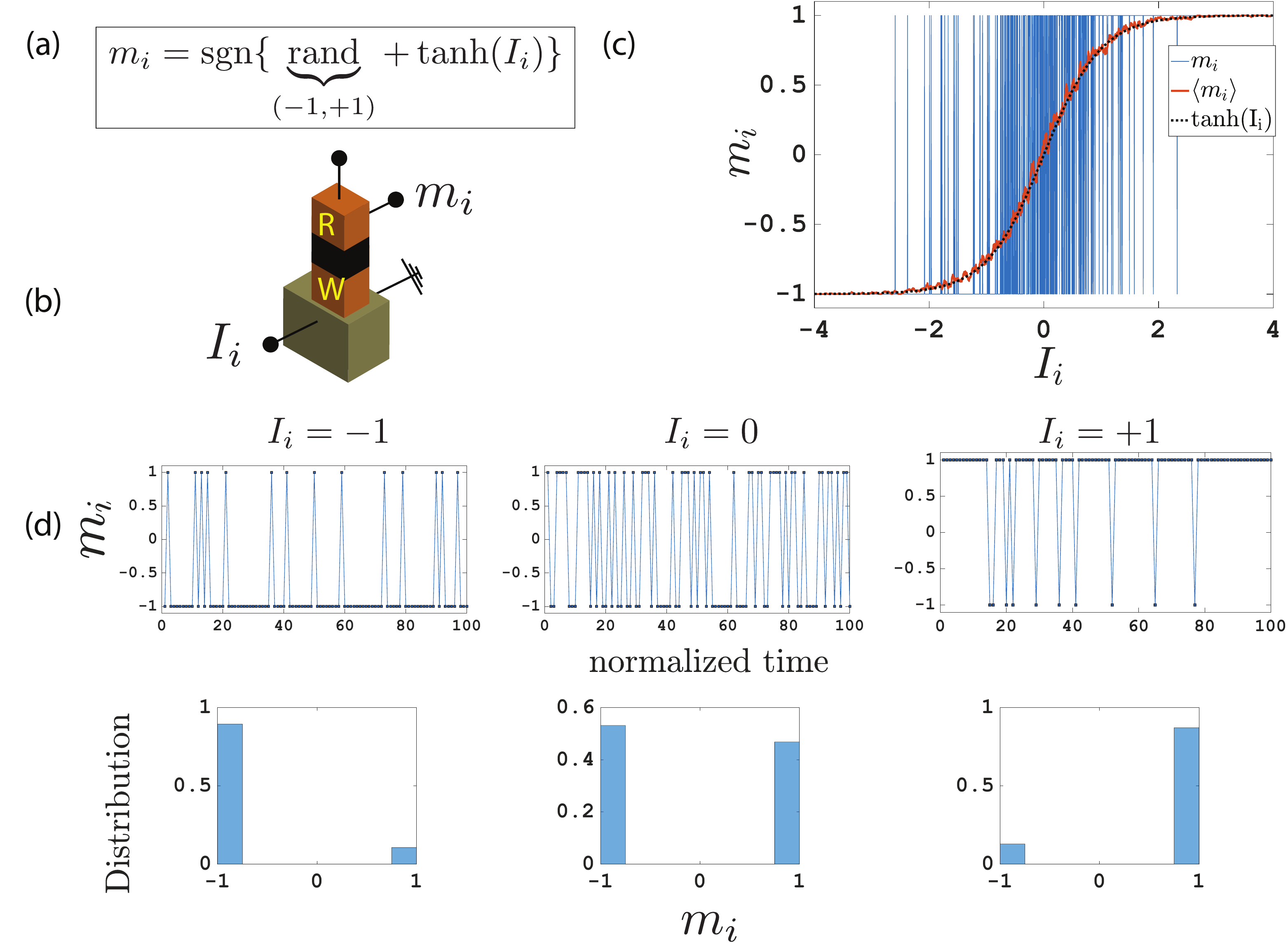}
\caption{\textbf{Generic building block for PSL:} (a) A generic model for PSL described by Eq.~(\ref{eq:sigmoid}) with distinct READ and WRITE units represented by the R/W icon shown in (b). Useful functionalities are obtained by interconnecting R/W units according to Eq.~(\ref{eq:weight}), ${I_i} = I_0 \times ({h_i} + \sum {{J_{ij}}{m_j}})$, with appropriately designed $\{h\}$ and $[J]$. (c) The blue trace shows the ``magnetization'' ($m_i$) obtained from Eq.~(\ref{eq:sigmoid}) as the current ($I_i$) is ramped. The red trace shows the sigmoid response obtained from an RC circuit which provides a moving average of the time-dependent ``magnetization'' which agrees very well with the black curve showing $\tanh(I_i)$. The bias terminal could involve a  voltage (V) instead of a current (I), just as the output could involve quantities other than magnetization. (d) The idealized telegraphic behavior of the model is shown at various bias points along with corresponding distributions. }
\label{fi:fig1}
\end{figure*}

Any random signal generator whose randomness can be tuned with a third terminal should be a suitable building block for PSL. The icon in Fig.~\ref{fi:fig1}b represents our generic building block whose input $I_i$ controls the output $m_i$ according to the equation (Fig.~\ref{fi:fig1}a),
\begin{equation}
{m_i}(t) = {\rm{sgn}}\{ \mathrm{rand(-1,1)} + \mathrm{tanh}({I_i}(t))\} 
\label{eq:sigmoid}
\end{equation}

\noindent where rand($-$1,+1) represents a random number uniformly distributed between $-$1 and +1. It is assumed to change every $\tau$ seconds which represents the retention time of individual p-bits. We normalize the time axis to $\tau$ so that t is  dimensionless and progresses in steps (0, 1, 2, $\ldots$). At each time step, if the input is zero, the output takes on a value of $-$1 or +1 with equal probability, as shown in the middle panel of Fig.~\ref{fi:fig1}d. A negative input $I_i$ makes negative values more likely (left panel) while a positive input makes positive values more likely (right panel). Fig.~1c shows $m_i(t)$ as the input is ramped from negative to positive values. Also shown is the time-averaged value of $m_i$ which equals $\mathrm{tanh}(I_i)$.

A possible physical implementation of p-bits could use stochastic nanomagnets with low energy barriers $\Delta$ whose retention time \cite{lopez2002transition}:
\begin{equation*}
\tau  = {\tau _0}\exp\left( {\Delta /kT} \right)
\end{equation*}
is very small, on the order of $\tau_0$  which is a material dependent quantity called the attempt time and is experimentally found to be $\approx 10 \rm \ ps - 1 \ ns$ \cite{lopez2002transition} among different magnetic materials. Such stochastic nanomagnets can be pinned to a given direction with spin currents that are at least an order of magnitude less than those needed to switch 40 kT magnets.  The sigmoidal tuning curve in Fig.~\ref{fi:fig1}c describing the time average of a fluctuating signal represents the essence of a p-bit. Purely CMOS implementations of a p-bit are possible \cite{palem2013ten,cheemalavagu2005probabilistic}, but the sigmoid seems like a natural feature of nanomagnets driven by spin currents. Indeed, the use of stochastic nanomagnets in the context of random number generators, stochastic oscillators and autonomous learning \cite{fukushima2014spin,choi2014magnetic,grollier2016spintronic} has been discussed in the literature. But performing ``invertible'' Boolean logic utilizing large scale correlations has not been discussed before to our knowledge.

Note that we are using the term \textit{invertibility} in the broader sense of relation inverses and not in the narrower sense of function inverses. For example, AND, when interpreted as a relation, consists of the set $\{\{1, 1 \rightarrow 1\}, \{0, 0 \rightarrow 0\}, \{1, 0 \rightarrow 0\}, \{0, 1 \rightarrow 0\}\}$ where each term is of the form $\{A, B\rightarrow \ \rm AND(A,B)\}$. The relation inverse of 0 is the set $\{\{0,
  0\}, \{0, 1\}, \{1, 0\}\}$  even though the corresponding functional inverse is not defined. What our scheme provides, probabilistically, is the relation inverse \cite{invertible_logic,invertible_logic2}. 

\begin{figure*}[t!]
\includegraphics[width=0.85\linewidth]{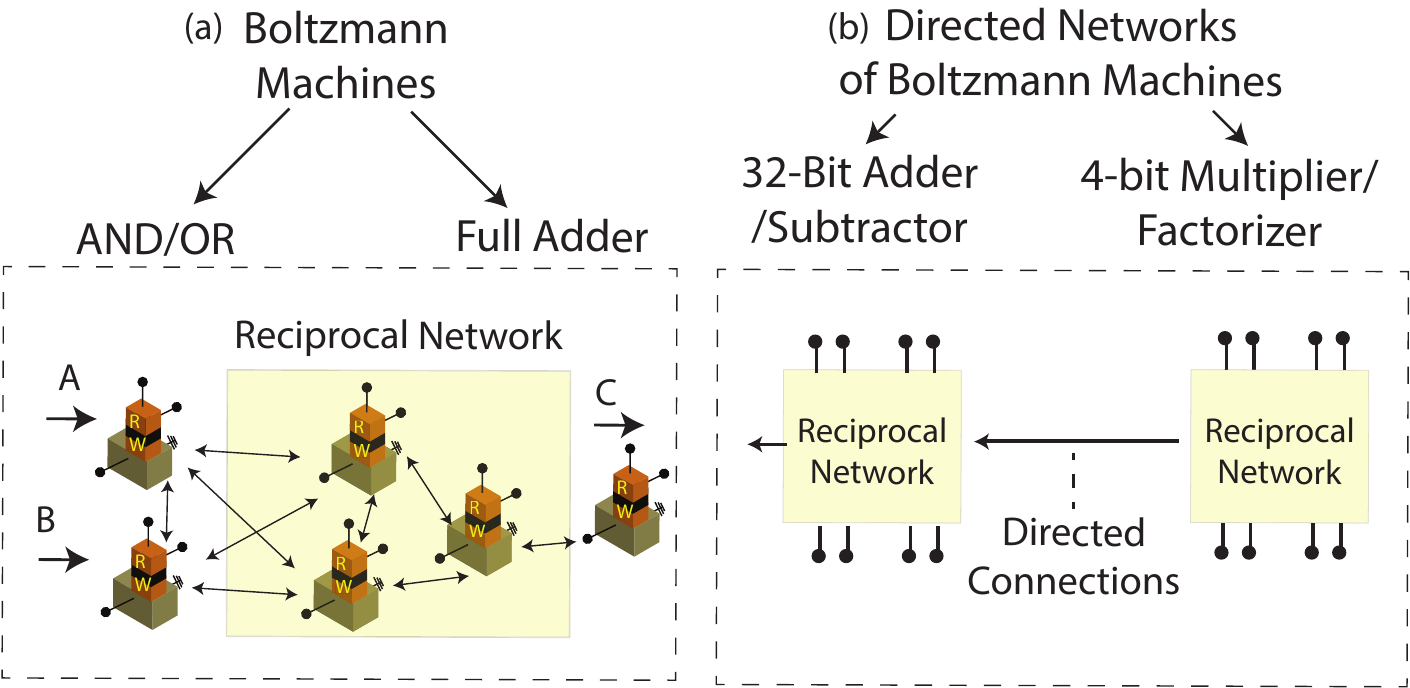}
\caption{\textbf{PSL designs discussed in this paper}: (a) Basic Boolean elements (AND/OR, Full Adder) are implemented as Boltzmann Machines based on  symmetrically coupled networks with $J_{ij}=J_{ji}$. (b) Complex Boolean functions like a 32-bit Ripple Carry Adder/Subtractor and 4-bit Multiplier/Factorizer are implemented by combining the reciprocal Boltzmann machines in a directed fashion.}
\label{fi:fig2}
\end{figure*}

\textit{Ensemble-average versus time-average:} A  sigmoidal response was presented in \cite{behin-aein_building_2016} for the ensemble-averaged magnetization of large barrier magnets biased  along a neutral state. This was proposed as a building block for both Ising computers as well as directed belief networks and a recent paper \cite{shim2016ising} describes a similar approach applied to a graph coloring problem. By contrast low barrier nanomagnets provide a sigmoidal response for the time-averaged magnetization and a suitably engineered network of such nanomagnets could cycle through the $2^N$ collective states at GHz rates, with an emphasis on the ``low energy states'' which can encode the solution to the combinatorial optimization problems, like the traveling salesman problem (TSP) as shown in \cite{sutton2016intrinsic}. Once the time-varying magnetization has been converted into a time-varying voltage through a READ circuit, a simple RC circuit can be used to extract the answer through a moving time average. For example, in Fig.~\ref{fi:fig1}c the red trace was obtained from the rapidly varying blue trace using an RC circuit in a SPICE simulation.

The central feature underlying both implementations is the \textit{p-bit} that acts like a tunable random number generator, providing an intrinsic sigmoidal response for the ensemble-averaged or the time-averaged magnetization as a function of the spin current. It is this response that allows us to \textit{correlate} the fluctuations of different p-bits in a useful manner by interconnecting them according to
\begin{equation}
{I_i}(t) = I_0 \times ({h_i}(t) + \sum_j {{J_{ij}}{m_j}(t)}) 
\label{eq:weight}
\end{equation}
where $h_i$ provides a local bias to magnet $i$ and $J_{ij}$ defines the effect of bit j to bit i, and $I_0$  sets a global scale for the strength of the interactions like an inverse ``pseudo-temperature'' giving a dimensionless current $I_i$ to each p-bit. The computation of $I_i(t)$ in terms of $m_j(t)$ in Eq.~(\ref{eq:weight})  is assumed instantaneous, in hardware implementations there can be interconnect delays that relate $m_j(t)$ to currents at a later time, $I_i (t')$. 

Equation (\ref{eq:sigmoid}) arises naturally from the physics of low barrier nanomagnets as we have discussed above. Equation (\ref{eq:weight}) represents the ``weight logic'' for which there are many candidates such as memristors \cite{yang2013memristive}, floating-gate based devices \cite{diep2014spin}, domain-wall based devices \cite{sengupta2016proposal}, standard CMOS \cite{yamaoka2016ising}. The suitability of these options will depend on the range of J values and the sparsity of the J-matrix.

Equations (\ref{eq:sigmoid}-\ref{eq:weight}) are essentially the same as the defining equations for Boltzmann machines introduced by Hinton and his collaborators \cite{ackley1985learning} which have had enormous impact in the field of machine learning, but they are usually implemented in software that is run on standard CMOS hardware. The primary contributions of this paper are threefold:
\begin{itemize}
\item \textit{Hardware implementation:}  It may seem ``obvious''  that an unstable magnet could provide a natural hardware for representing a p-bit, but we would like to stress a less obvious point. To the best of our knowledge, simple two-terminal devices are not suitable for constructing large scale correlated networks of the type envisioned here. Instead, we need three-terminal building blocks with transistor-like gain and input-output isolation as shown in Fig.~\ref{fi:fig1}b  \cite{behin-aein_building_2016}. To stress this point, we describe a concrete implementation of a Boolean function using detailed nanomagnet and transport simulations that are in good agreement with those obtained by the generic model based on Eq.~(\ref{eq:sigmoid}). All other results in this paper are based on Eq.~(\ref{eq:sigmoid}) in order to emphasize the generality of the concept of p-bits which need not necessarily be nanomagnet-based \cite{yamaoka201524,inagaki2016large}.

\item \textit{Boltzmann machines (BM) for invertible Boolean logic (Fig.2a):} Much of the current emphasis on BMs is on ``learning'' giving rise to the concept of restricted Boltzmann machines \cite{salakhutdinov2007restricted}. By contrast this paper is about Boolean logic, extending an established method for Hopfield networks \cite{amit1992modeling} to provide a mathematical prescription to turn \textit{any} Boolean truth table into a symmetric J-matrix (Eq.~(\ref{eq:weight}), with $J_{ij}=J_{ji}$), in one shot with no ``learning''  being involved. This design principle seems quite robust, functioning satisfactorily even when the J-matrix elements are rounded off, so that the required interconnections are relatively sparse and quantized which simplifies the hardware implementation. The numerical probabilities agree well with those predicted from the energy functional.
\begin{equation}
E(\{m\}) =- I_0 \times \bigg(\sum_{i,j} \frac{1}{2} \left(J_{ij} m_i m_j\right) + \sum_i h_{i} m_i\bigg)
\end{equation} 

using the Boltzmann law:

\begin{equation}
P(\{m\}) = \frac{\exp(-E)}{ \sum_{i,j} \exp(-E)}
\label{eq:BL}
\end{equation}

Most importantly we show that \textit{the resulting Boolean gates are invertible:} not only  do they provide the correct output for a given input, for a given output they provide the correct input(s). If the given output is consistent with multiple inputs, the system fluctuates among all possible answers. This remarkable property of invertibility is absent in standard digital circuits and could help provide solutions to the Boolean satisfiability problem (Fig.~\ref{fi:figure8}) \cite{du1997satisfiability}. 

\item \textit{Directed networks of BM (Fig.2b):} Finally we show that individual BM's can be connected to perform \textit{precise} arithmetic operations which are the norm in standard digital logic, but quite surprising for BM which are more like a collection of interacting particles than like a digital circuit. We show that a 32-bit adder converges to the one correct sum  out of $2^{33} \approx 8$ billion possibilities when the interaction parameter is suddenly turned up from say $I_0= 0.25$ to $I_0=5$. This can be likened to quenching a molten liquid and \textit{getting a perfect crystal.} What we expect is plenty of defects, distributed differently everytime we do the experiment. That is exactly what we get if the individual BM Full adders comprising the 32-bit adder are connected bidirectionally ($J_{ij}=J_{ji}$). But by making the connection between Adders directed ($J_{ij} \neq J_{ji}$), we obtain the striking accuracy of digital circuits while largely retaining the invertibility of BM.  This is a key result that we establish with extensive examples including a 4-multiplier which in inverted mode functions as a factorizer. 

\end{itemize}

\noindent Each of these three contributions is described in detail in the three sections that follow.
  \vspace{0em}
\section{An example hardware Implementation of PSL}
  \vspace{0em}
\label{sec:hw}
\begin{figure}
\includegraphics[width=\linewidth]{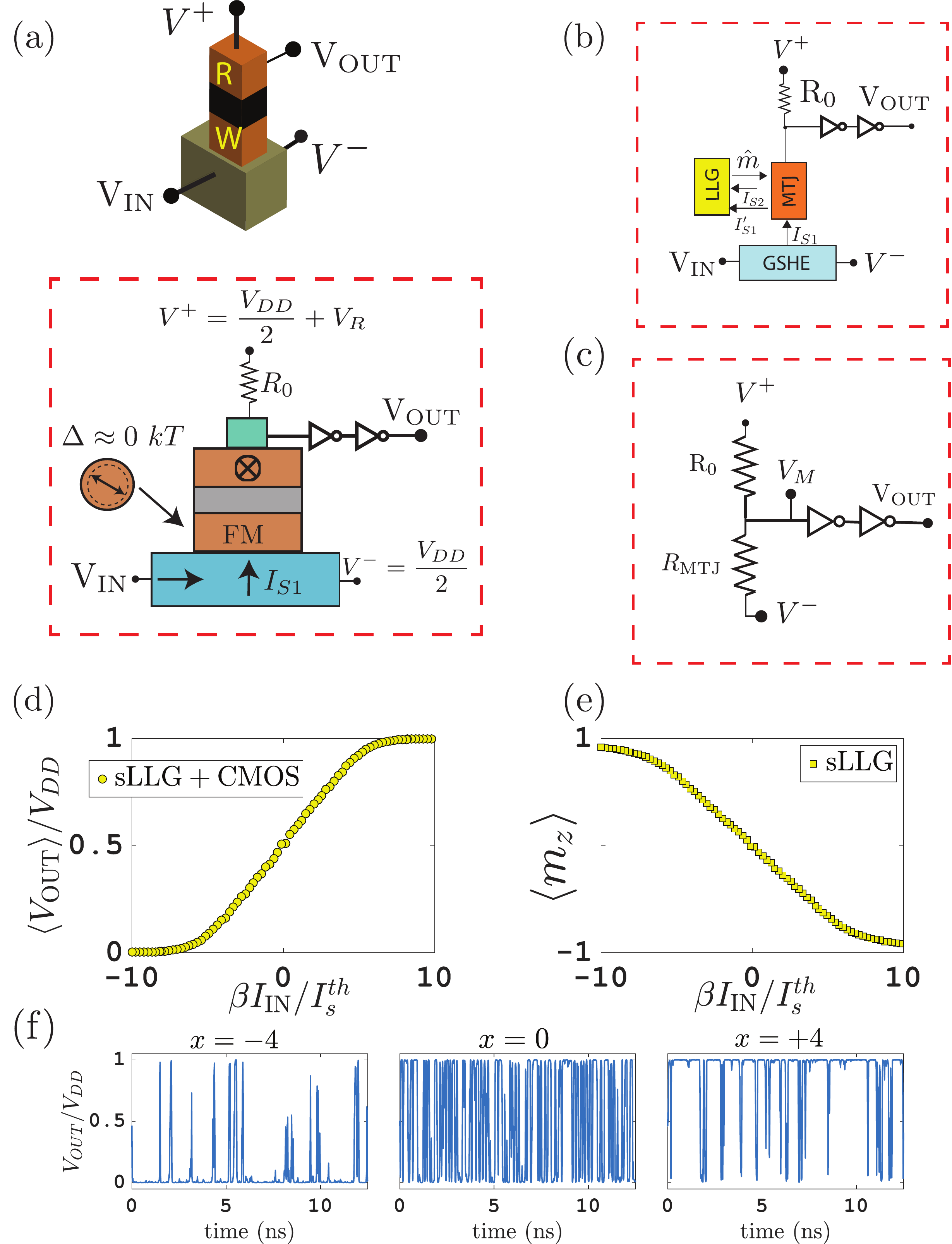}
\caption{\textbf{CMOS-assisted implementation of p-bits:} (a) A possible CMOS-assisted implementation of p-bits that have a separate READ/WRITE paths. A GSHE layer provides a spin current that pins the magnetization of circular magnets ($\Delta \approx 0 \ kT)$. The change in magnetization is sensed by an MTJ and amplified by two CMOS inverters that act as a buffer, providing the necessary isolation and gain. (b) Self-consistent, modular modeling of transport and magnetization dynamics. See ``Assumptions of the model'' in the text. (c) Equivalent READ circuit. (d) SPICE-based average output voltage normalized to the $V_{DD}=0.8 \rm \ V$ of 14 nm FinFET HP-inverters \cite{predictive_tech}. (e) sLLG-based average magnetization of the circular magnet as a function of the spin current (averaged over $500\  \rm ns$ for each bias point with a time step of $\Delta t=0.05 \rm \ ps$, 10 million points per marker), normalized to the GSHE gain and the thermal noise strength, $I_s^{th}$. (f) The time-dependent output voltage at various bias points.}
\label{fi:figure3}
\end{figure}

\begin{figure}
\includegraphics[width=0.95\linewidth]{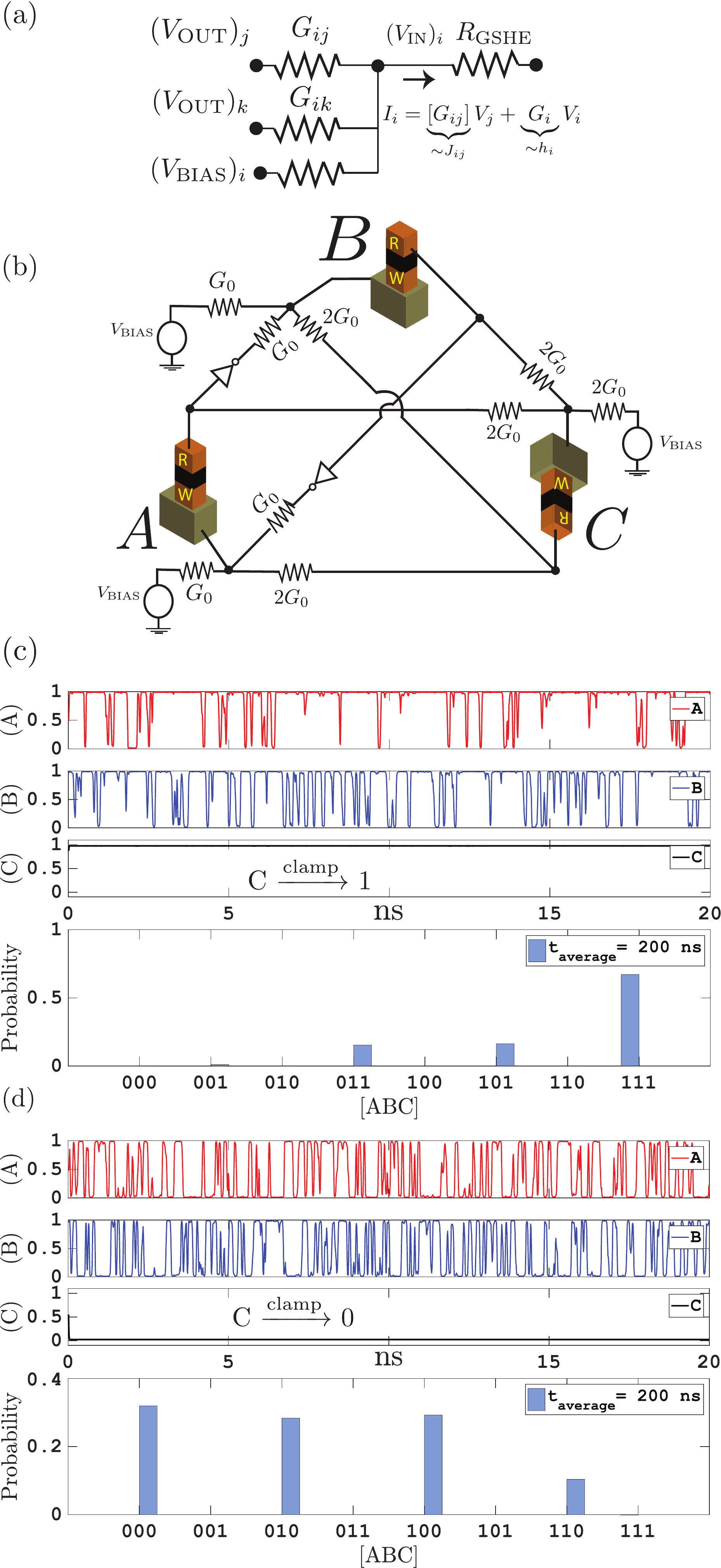}
\caption{\textbf{An invertible AND gate:} (a) Passive resistor network that is used to obtain the connection terms $J_{ij}$ to correlate p-bits. The output impedance $R_{ij}=1/G_{ij}$ is much smaller than the input impedance $R_{GSHE}$, allowing separate voltages to add at the input of the $\rm i^{th}$ p-bit. (b) Explicit implementation of an AND gate based on Eq.~(\ref{eq:and}). (c) When C is clamped to 1, A and B spend most of their time in the (11) state, the only combination consistent with C=1. (d) The invertible operation of the AND gate when the C gate is clamped to a zero, while A and B are left floating. A and B bits fluctuate between 3 possible combinations consistent with C=0, (A,B)=(00),(01),(10). The time response of A,B,C voltages are normalized by $V_{DD}$. Histogram is obtained by averaging over 200 ns of thresholded voltages, only the first 20 ns of A,B,C voltages are shown for clarity.}
\label{fi:figure4}
\end{figure}

To ensure that individual p-bits can be interconnected to produce robust correlations, it is important to have separate terminals for writing (more correctly biasing) and reading, marked W and R respectively in Fig.~\ref{fi:figure3}a. With IMA nanomagnets (e.g circular nanomagnets) this could be accomplished following existing experiments \cite{liu2012spin,locatelli2014noise} using the giant spin Hall effect (GSHE). Recent experiments using a built-in exchange bias \cite{van2016field,lau2016spin,smith2016external,fukami2016magnetization} could make this approach applicable to PMA as well. Note however, that these experiments have all been performed with stable free layers, and would have to be carried out with low barrier magnets in order to establish their suitability for the implementation of p-bits. As the field progresses, one can expect the bias terminal to involve voltage control \cite{heron2014deterministic,manipatruni2015spin} instead of current control, just as the output could involve quantities other than magnetization. We will now show a concrete implementation of a Boolean function using minimal CMOS circuitry in conjunction with stochastic nanomagnets through detailed nanomagnet and transport simulations that are in good agreement with those obtained from the generic model based on Eq.~(\ref{eq:sigmoid}).

Fig.~\ref{fi:figure3}a shows a possible, CMOS-assisted p-bit that has a separate READ and WRITE path. The device consists of a heavy metal exhibiting Giant Spin Hall Effect (GSHE) that drives a circular magnet which replaces the usual elliptical magnets in order to provide the stochasticity needed for the magnetization. A small read current, which is assumed to not disturb the magnetization of the free layer in our design, that flows through the fixed layer is used to sense the instantaneous magnetization, which is amplified and isolated by two inverters that act as a buffer. This structure is very similar to the experimentally demonstrated  GSHE switching of elliptical magnets that were similarly read-out by an MTJ \cite{liu2012spin}, with the only exception that the elliptical magnets are replaced by circular magnets with an aspect ratio of one. This device could be viewed as replacing the free layers of the GSHE-driven MTJs demonstrated in \cite{liu2012spin} with those in the telegraphic regime \cite{koch2000thermally,urazhdin2003current,krivorotov2004temperature,locatelli2014noise} . 

 In the presence of thermal noise the magnetization of such a circular magnet rotates in the plane of the circle without a preferred easy-axis that that would have arisen due to the shape anisotropy, effectively making its thermal stability $\Delta\approx 0 \ \rm kT$ \cite{khvalkovskiy2013basic}. This magnetization can be pinned by a spin current that is generated by flowing a charge current through the GSHE layer. The magnetic field driven sigmoidal responses of magnetization for such circular magnets have experimentally been demonstrated \cite{cowburn2000property,debashis2016experimental}, while the spin current driven pinning has not been demonstrated to our knowledge. Using validated modules for transport and magnetization dynamics \cite{camsari2015modular} (Fig.~\ref{fi:figure3}b), we solve the stochastic Landau-Lifshitz-Gilbert (sLLG) equation in the presence of thermal noise and a GSHE current. The following subsection shows detailed simulation parameters. 
 
\textit{Sigmoidal response:} A long-time average ($t=500\rm \ ns)$ of the magnetization $\langle m_z \rangle $ as a function of a GSHE-generated spin current is plotted in Fig.~\ref{fi:figure3}e that displays the desired sigmoidal characteristic for p-bits dictated by Eq.~(\ref{eq:sigmoid}). The x-axis of Fig.~\ref{fi:figure3}e is normalized to the geometric gain factor that relates the charge current to the spin current exerted \cite{PhysRevLett.106.036601,hong2016spin}:
 \begin{equation}
 \beta \equiv \frac{I_s}{I_c}=\theta_{\rm SH} \frac{L_{FM}}{t}\bigg(1-\mathrm{sech}\left(\frac{t}{\lambda}\right)\bigg)
 \label{eq:gshe}
\end{equation}
where $\theta_{SH}$ is the Hall angle, $ t $ is the thickness and $\lambda$ is the spin-relaxation length of the heavy metal. The quantity $\beta$ can be made to be much greater than 1 providing an intrinsic gain \cite{datta2012non},  however for the parameters used in the present examples, $\beta$ is $\approx 1.5$. 

Another quantity that is used to normalize the x-axis of Fig.~\ref{fi:figure3}e is the ``thermal spin current'' that corresponds to the strength of the thermal noise that needs to be overcome for a circular magnet to be pinned in a given direction: 
\begin{equation}
I_s^{th}  = \bigg(\frac{4q}{\hbar}\bigg) \alpha \big(kT\big)
\label{eq:is_th}
\end{equation}
where q is electron charge, $\alpha$ is the damping coefficient of the magnet. $I_s^{th}$,  $I_s$ and $I_c$ all have units of charge current, therefore we can define the dimensionless interaction parameter, $I_0$ of Eq.~\ref{eq:weight} as  $I_0 \equiv \beta I_c  / I_s^{th}  = I_s / I_s^{th}$. 

It can be seen from Fig.~\ref{fi:figure3}e that when the applied spin current  $\beta I_c  / I_s^{th}  = I_s / I_s^{th}\approx 10$, the magnetization of the circular magnet is pinned in the $\pm z$ directions for these particular parameters. For PMA magnets with low barriers ($\Delta \ll  kT$), the pinning current is independent of the volume as long as increasing the volume does not invalidate the $\Delta \ll  kT$ assumption. This can be analytically shown from a 1D Fokker-Planck equation \cite{butler2012switching}, and we have reproduced this behavior directly from sLLG simulations.  For the in-plane (circular) magnets considered here, the pinning current in general has a $M_s$ and $\rm Vol.$ dependence and the dimensionless pinning current can be larger. 

Nevertheless, it is possible to estimate the thermal spin current for typical damping coefficients of $\alpha=0.01-0.1$, $I_s^{th}$ is $\approx 0.25 \  \mu A - 2.5 \  \mu A$.  Pinning currents for superparamagnets are at least an order of magnitude smaller than the critical switching currents of stable magnets \cite{kent2015new}. $I_s^{th}$, defined by Eq.~(\ref{eq:is_th}) also sets the scale for $I_0$ defined in Eq.~(\ref{eq:weight}) suggesting that a stochastic nanomagnet based implementation of PSL could be more energy efficient than the standard spin-torque switching of stable magnets that suffer from high current densities.

\textit{Need for three-terminal devices with READ-WRITE separation:} Note that a crucial function of the READ circuit and the CMOS transistors in this design is the ability to turn the magnetization into an output voltage that is proportional to $m_z$,  providing gain for fan-out and isolation to avoid any read disturb. Indeed, a critical requirement for any other alternative implementations of p-bits is the need for three terminal devices with separate READ and WRITE paths to provide gain and isolation. In this particular design these features come in by directly integrating CMOS transistors, but CMOS-free, all-magnetic designs with these characteristics have been proposed \cite{datta2012non,morris2012mlogic}. Our purpose is to simply show how a p-bit can be realized by using experimentally demonstrated technology. Alternative designs are beyond the scope of this paper. 

\textit{READ Circuit:} For the output  to provide symmetric voltage swings on the GSHE layer, the minus supply $V^-$ needs to be set to $V_{DD}/2$ since $V_{\rm OUT}$ ranges between 0 and $V_{DD}$.  $V^+$ is set to $V_{DD}/2+V_{R}$ where $V_R$ is a small READ voltage that is amplified by the inverters.  We assume a simple, bias-independent MTJ model \cite{datta2012voltage}:
\begin{equation}
G_{\mathrm{MTJ}} =  G_0(1+P^2 m_z),
\label{eq:mtj}
\end{equation}
where P is the interface polarization and $G_0$ is the average MTJ conductance. Setting the reference resistance (Fig.~(\ref{fi:figure3}c)) $R_0$ equal to $G_0^{-1}$, the input voltage to the inverters, $V_M$ in FIG.~(\ref{fi:fig2}d) becomes:
 \begin{equation}
 V_M = \frac{V_{DD}}{2}+ \frac{V_R}{2+m_z P^2}
 \label{eq:vm}
 \end{equation}
In the absence of a bias  $\langle m_z \rangle$ becomes $0$ and the middle voltage fluctuates around the mean  $\langle V_M \rangle = V_{DD}/2+V_R/2$.  This requires the inverter characteristic to be shifted to this value to produce a telegraphic output that fluctuates between 0 and $V_{DD}$ with equal probability (Fig.~\ref{fi:figure3}f). This shift is easily engineered by sizing the pFET and nFET transistors differently, a wider pFET shifts the inverter characteristic towards $V_{DD}$, as we will show in the next subsection. 

\textit{Interconnection matrix:} A passive resistor network can be used as a possible interconnection scheme to correlate the p-bits as shown in Fig.~\ref{fi:figure4}. A proper design of the interconnection matrix J that has only a few discrete values ensures  a minimal number of different conductances ($G_{ij}$). In this demonstrated example the AND gate requires only 2 unique, discrete conductance values.

The spin currents that need to be delivered to each p-bit are on the order of a few $\mu A$ and can be generated with charge currents that are even smaller, due to the GSHE gain. This means the interconnection resistances $R_{ij}$ could be on the order of  100 k$\Omega$'s since the voltage drops across these resistances are around $V_{\rm OUT} - V^-\approx \pm 0.5 $ V. Since the GSHE ground $V^-=V_{DD}/2$ simply shifts all the voltages to get symmetric $\pm$  swings, we define the voltages $(V'_{\mathrm{OUT}})_i= (V_{\mathrm{OUT}})_i-V^-$. Then input currents to each p-bit can be expressed (Fig.~\ref{fi:figure4}a):
\begin{equation}
\big(I_{\mathrm{IN}}\big)_i = \sum_j  G_{ij} (V'_{\mathrm{OUT}})+ G_i  (V'_{\mathrm{BIAS}})
\label{eq:weight_int}
\end{equation}
assuming $\sum_j G_{ij} \ll G_{\mathrm GSHE}$ since  the heavy metal resistances are typically much less than hundreds of k$\Omega$. We have verified the validity of Eq.~(\ref{eq:weight_int}) by SPICE simulations, for the parameters chosen for these examples. 

As a result, we observe that Eq.~(\ref{eq:weight_int}) constitutes a hardware mapping for the interconnections of  Eq.~(\ref{eq:weight}). In this scheme $G_{ij}$ conductances are initially adjusted to obtain a global interaction strength $I_0$ for a given problem. Alternatively, the interaction strength can be adjusted electrically by varying the supply voltages. 

\textit{Invertible AND Gate:} Fig.~\ref{fi:figure4}b shows an explicit implementation of an invertible AND gate ($A\cap B=C)$ corresponding to [J] and $\{h\}$ matrices \cite{biamonte2008nonperturbative} that have 3 unique, integer entries: 
\begin{equation}
J = \bbordermatrix{~ & \rm A & \rm B & \rm C \cr
              \rm A & 0 & -1 & +2 \cr
              \rm B & -1  & 0 & +2 \cr
              \rm C & +2 & +2 &  0 \cr } \quad h^T=\begin{bmatrix} +1 & +1 & -2 \end{bmatrix}
             \label{eq:and}
\end{equation}

In Fig.~\ref{fi:figure4}d, we show the \textit{inverse} operation of the AND gate where we clamp the output bit C to a 0 or 1 by  the bias voltage attached  to its input terminal. The interconnection resistance is chosen to be $R_0=125 \ \rm k\Omega$ that roughly provides $\approx \pm6\  \mu$A of charge current to each p-bit, corresponding to an $I_0 \approx 3.5$ for the chosen parameters. 

\textit{Generating the histogram:} At the end of the simulation (t=200 ns), we threshold the voltage output of A,B and C by legislating all voltages above $V_{DD}/2 = 0.4 \ \rm V$ to be  1, and below $V_{DD}/2$ to be 0. Then a histogram output for the thresholded word [ABC] is obtained and normalized to unit probability. Clamping the output to 0 and letting A and B float,  make A and B fluctuate in a correlated manner and they visit the three possible states (00, 01, 10) with approximately equal probability. Resolving the output 0 to the three possible input combinations is, in a way ``factorizing'' the output.  Conversely, clamping the output to 1 produces a strong (11) peak in the histogram of [ABC], which is the only consistent input combination for C=1 (Fig.~\ref{fi:figure4}c-d). 

\textit{Assumptions of the model:} We have made several simplifying assumptions while modeling the hardware implementation of a p-bit. (1) The READ voltage that is amplified by the inverters produces a small current that passes through the circular magnet and might potentially disturb its current state. We assumed that this current (labeled as $I_{S2}$ in Fig.~\ref{fi:figure3}b) is negligible and do not affect the magnetization of the stochastic magnet. (2) We assumed that the spin current generated by the heavy metal is deposited to the free layer with perfect efficiency ($I'_{S1}=I_{S1}$ in Fig.~\ref{fi:figure3}b), however, depending on the interface properties this conversion factor can be less  than 100$\%$.  (3) We have also assumed that the fixed layer does not produce a notable stray field on the circular magnet. Note that the presence of such a constant field would simply shift the sigmoidal behavior presented in Fig.~\ref{fi:figure3}d-e to the right (or left) and could have been offset by a constant bias current. (4) Finally, we have neglected the resistance of the GSHE portion in the READ circuit (Fig.~\ref{fi:figure3}c), assuming the MTJ resistance would be dominant in this path. 
\subsection{Detailed Simulation Parameters}
This section shows the details of simulation parameters for the hardware implementation of p-bits that are used for Fig.~\ref{fi:figure3}$-$\ref{fi:figure4}. 

\textit{sLLG for stochastic circular magnets:} The magnetization of a circular nanomagnet described as $\hat m_i$ is obtained from the stochastic Landau-Lifshitz-Gilbert (sLLG)
equation:
\begin{subequations}
\begin{align}
&(1+\alpha^2)\frac{d\hat m_i}{dt} = -|\gamma|{\hat m_i \times \vec{H}_i} - \alpha |\gamma| (\hat m_i \times \hat m_i \times \vec{H}_i)\nonumber \\ &+  \frac{1}{q  N_i}(\hat m_i \times \vec{I}_{Si} \times \hat m_i)  + \left(\frac{\alpha}{q N_i} (\hat m_i \times \vec{I}_{Si})\right)
\label{sLLG}
\end{align}
\end{subequations}
where $\alpha$ is the damping coefficient, q is the electron charge, $\gamma$ is the electron gyromagnetic ratio, $I_s$ is the spin current that is assumed to be uniformly distributed over the total number of spins in the macrospin, $N_i = M_s \mathrm{Vol.}/\mu_B$, $\mu_B$ being the Bohr magneton. It is assumed that the spin current generated from the GSHE layer is polarized in the z-direction, such that $\vec{I}_{Si} = I_S \hat z$. $\vec{H}_i$ is the effective field of the circular magnet, where the uniaxial anisotropy is assumed to be negligible, but there is still a strong demagnetizing field. The thermal fluctuations also enter through the effective magnetic field: $\vec{H}_i  = -4\pi M_s m_x \hat x + \vec{H}_{th}$, $x$-axis being the out-of-plane direction of the magnet, and $\langle |\vec{H}_{th}|^2 \rangle = 2  \alpha  kT / (|\gamma| M_s \mathrm{Vol.})$ in units [$\rm Oe^2 / Hz$] with zero mean, and equal in all three directions. Table~\ref{table:1} shows the parameters used in Figs.~\ref{fi:figure3}$-$\ref{fi:figure4}. We note that this parameter selection is simply one possibility, many other parameters could have been used with no change in the basic conclusions. 

\begin{table}[b!]
\renewcommand{\arraystretch}{1.65}
\centering
\begin{tabular}{c c}
\hline
\bfseries Parameters & \bfseries Value\\
\hline
Saturation magnetization ($M_s  $) & 300 \  emu/cc \\
Magnet diameter ($\Phi$), thickness (t)  & 15 nm, 0.5 nm \\
MTJ Polarization (P) (Eq.~(\ref{eq:mtj}))  & 0.5 \\ 
MTJ Conductance ($G_0$) (Eq.~(\ref{eq:mtj}))   &  176 $\mu$S  \\ 
Damping coefficient ($\alpha$) & 0.1 \\
Spin Hall Length, Width (Eq.~(\ref{eq:gshe}))  & $L=W= 15$ nm \\
Hall Angle, Spin relax. length &\hspace{-10pt} $\theta$=0.5 \cite{demasius2016enhanced}, $\lambda_{sf}=$2.1 \rm nm\cite{pai2012spin}\\ 
Spin Hall res. $(\rho)$, thickness ($t$) & 200 $\mu\Omega$-cm \cite{hao2015giant}, 3.15 nm \\
Temperature ($T$) & 300 \rm \ K \\
CMOS Models & 14nm  HP-FinFET \cite{predictive_tech} \\
Supply and READ Voltage &\hspace{-15pt}  $V_{DD}=0.8\rm \ V$, $V_R=0.5\rm\ V$ \\
Timestep for  transient sim. (SPICE)   &$\Delta$t = 0.05 \ ps\\
\hline \hline
\end{tabular}
\caption{Parameters used for simulations in Figs.~\ref{fi:figure3}$-$\ref{fi:figure4}.}
\label{table:1}
\end{table}

\textit{Obtaining the sigmoidal response of CMOS+sLLG:} Each data point in the sigmoids shown in  Figs.~\ref{fi:figure3}$-$\ref{fi:figure4}  is obtained by averaging the z-component of the magnetization after 500 ns, with a time-step of $\Delta t = 0.05 \ \rm ps$. The CMOS inverter characterestics in conjunction with a spherical representation-based sLLG are obtained using the modular framework developed in \cite{camsari2015modular}  using HSPICE. 

\begin{figure}[t!]
\includegraphics[width=1\linewidth]{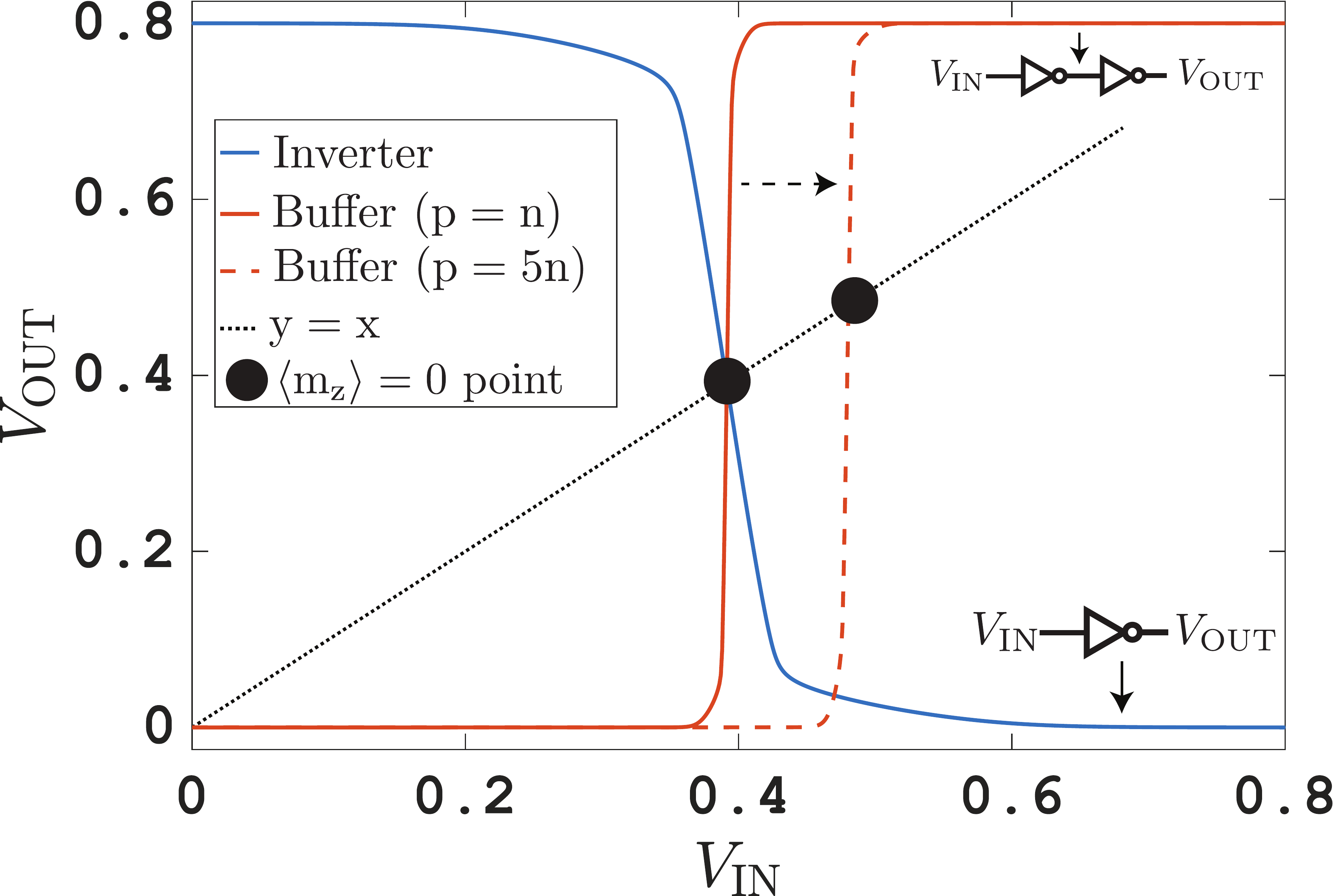}
\caption{\textbf{14 nm PTM, Inverter/Buffer:} DC response of 14 nm high performance (HP) FinFETs based on \cite{predictive_tech} for an inverter and buffer. Sizing the transistors differently allows the switching point to be shifted.}
\label{fi:figure5}
\end{figure}

\textit{14 nm FinFET Inverter Characteristics:} Fig.~\ref{fi:figure5} shows the input/output characteristics of the single and double inverters that are used to amplify the  stochastic signal that is generated by the MTJ (Fig.~\ref{fi:figure3}). At zero-bias from the GSHE, the amplified signal $V_M$ (Eq.~\ref{eq:vm}) is in the middle of $V^+$ and $V^-$ which is $V_{DD}/2+V_R/2$. The buffer response can be shifted to this value by increasing the size of pFETs, as shown in Fig.~\ref{fi:figure5}.

\section{Invertible Boolean logic with Boltzmann Machines}
  \vspace{0em}
\label{sec:truth}

\begin{figure}[t!]
\includegraphics[width=\linewidth]{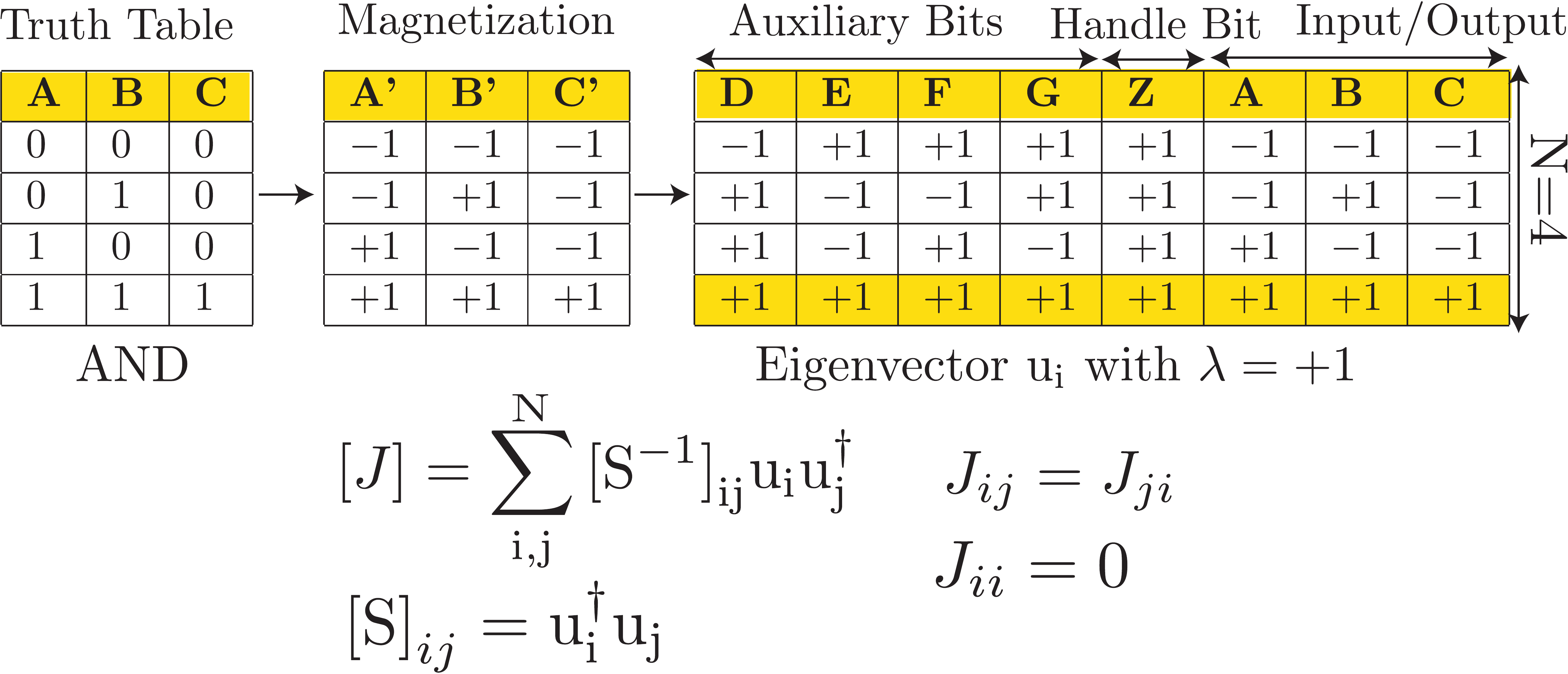}
\caption{\textbf{Truth Table to J-Matrix:} A given truth table is first transformed from binary to bipolar variables by using the transformation $m=2t - 1 $, where m and t represent the magnetization and binary values of the truth table. Additional bits are introduced to each line of the truth table to ensure that the resultant S-matrix is invertible. The indices $\rm i,j$ correspond to the number of lines in the truth table. $\rm u_i, u_j$ are column vectors. As an example, we have shown auxiliary bits that result in an S-matrix equal to the identity matrix, since the eigenvectors are orthogonal.  The J-matrix is then obtained by Eq.~(\ref{eq:amit}) which ensures that the truth table corresponds to the low energy states of the Boltzmann machines according to Eq.~(\ref{eq:BL}). A handle bit  of +1 is introduced to each line of the truth table which can be biased to ensure  that the complementary truth table does not appear along with the desired one. This bit also allows a truth table to be electrically reconfigured into its complement.}
\label{fi:figure6}
\end{figure}

We now present a mathematical prescription that shows how any given truth table can be implemented in terms of Boltzmann Machines, in ``one shot'' with no learning being involved, unlike much of the past work in this area (See for example, \cite{Sejnowski1986learning,patarnello1987learning}). In Section~\ref{sec:hw}, we chose a simple [J] and $\{h\}$ matrix to implement an AND gate based on \cite{biamonte2008nonperturbative}. In this section, we outline a general approach to show  how any truth table can be implemented in terms of such matrices. Our approach, pictorially described in Fig.~\ref{fi:figure6}, begins by transforming a given truth table from binary $(0,1)$ to  bipolar $(-1,+1)$ variables. The lines of the truth table are then required to be eigenvectors each with eigenvalue +1, all other eigenvectors are assumed to have eigenvalues equal to 0. This leads to the following prescription for J as shown in Fig.~\ref{fi:figure6}:

\begin{subequations}
\begin{align}
[J ] = \sum_{i,j}\mathrm{[S^{-1}]_{ij} u_i u_j^\dagger}\label{eq:amit}
\\
\rm S_{ij}   = u_i^\dagger u_j 
\end{align}
\end{subequations}

\noindent where $u_i$ are the eigenvectors corresponding to lines in the truth table of a Boolean operation and S is a projection matrix that accounts for the non-orthogonality of the vectors defined by different lines of the truth table. Note that the resultant J-matrix is  always symmetric ($J_{ij}=J_{ji}$) with diagonal terms that are subtracted in our models such that $J_{ii}=0$. The number of p-bits in the system is made greater than the number of lines in a truth table through the addition of hidden units (Fig.~\ref{fi:figure6}) to ensure that the number of conditions we impose is less than the dimension of the space defined by the number of p-bits. 

Another important aspect in the construction of [J] is that an eigenvector $\rm u_i$ implies that its complement $\rm -u_i$ is also a valid eigenvector. However only one of these might belong to a truth table. We introduce a ``handle'' bit to each $\rm u_i$ that is biased $(h_i)$ to distinguish  complementary eigenvectors. These handle bits provide the added benefit of reconfigurability. For example, AND and OR gates have complementary truth tables, and a given gate can be electrically reconfigured as an AND or an OR gate using the handle bit.

  \vspace{0em}

\textit{J-Matrices for AND/FA}: We now provide the details of the J-matrix for the AND gate, obtained using the prescription shown in Fig.~\ref{fi:figure6} based on Eq.~(\ref{eq:amit}). The eigenvectors of the truth table for the AND in Fig.~\ref{fi:figure6} are placed into a matrix U, such that $\rm U = [ u_1 \ u_2  \ u_3 \  u_4]$, where $u_1$ is the first row of the matrix shown in Fig.~\ref{fi:figure6},  $ u_1 = [-1 +1 +1 +1 +1 -1 -1 -1]^T $ and so on. In matrix notation, the S-matrix can be written as:
\begin{equation}
S= U^T U  = 8 \ I_{4\times 4} 
\end{equation}
Then the J-matrix becomes:
\begin{equation}
J = \sum_{ij} \underbrace{[S^{-1}]_{ij}}_{1/8 \ \delta_{ij}} u_i u_j^\dagger = 1/8 \sum_i  u_i u_i^\dagger
\end{equation}
\begin{figure*}[t!]
\includegraphics[width=0.75\linewidth]{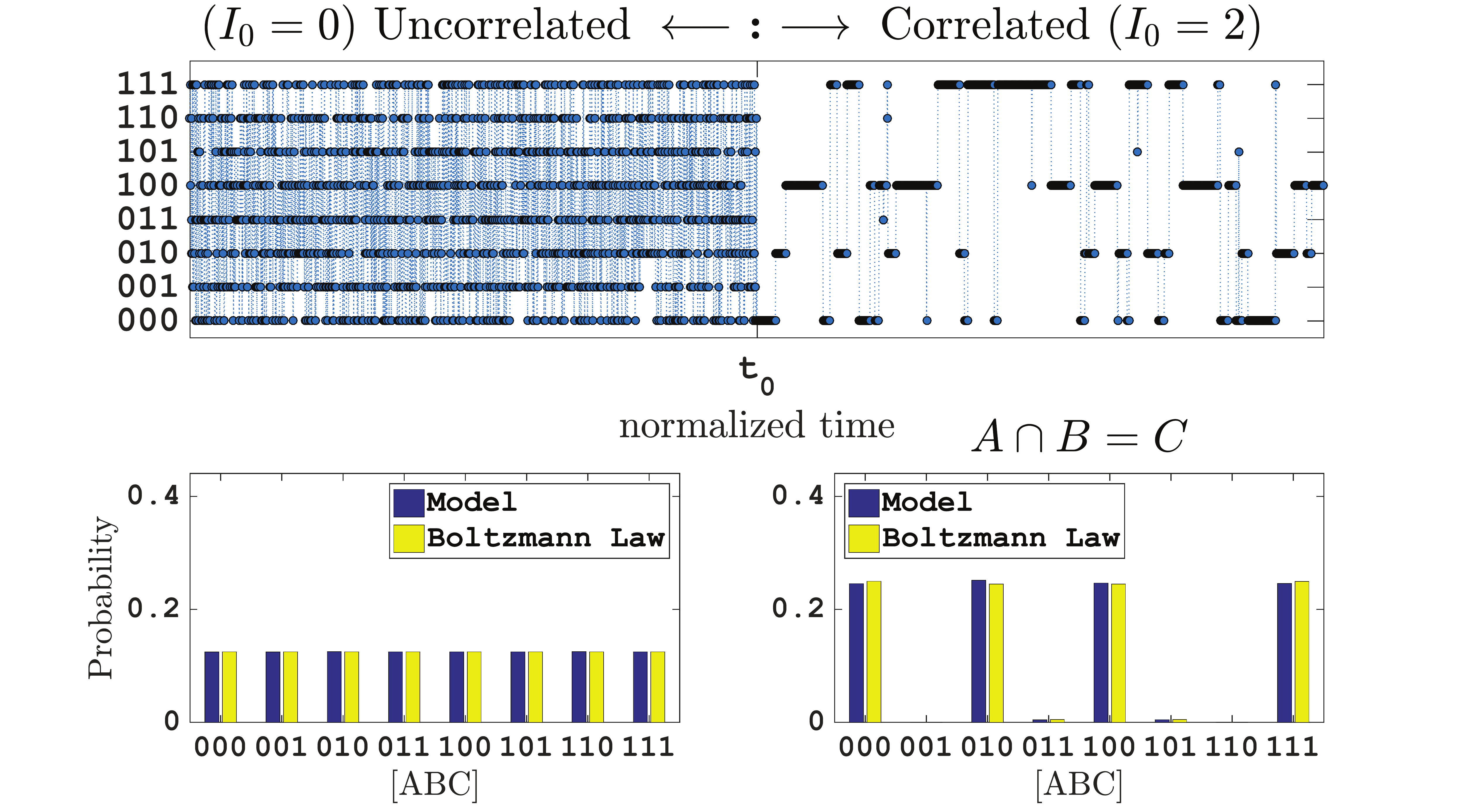}
\caption{\textbf{Correlated p-bits, AND Gate}: When the interaction strength ($I_0$) is zero, p-bits produce uncorrelated noise, visiting all possible states with equal probability. In this example, the interaction strength (pseudo inverse-temperature) is suddenly increased from 0 to 2 as a step function at $t=t_0$, to effectively ``quench''  the network. This correlates the p-bits to produce the truth table of an AND gate (AND: $A \cap B = C$). Note that after this quenching, the p-bits only visit the low energy states corresponding to the truth table of the AND gate and once the system is in one of the low energy states, it tends to stay there for a while, until being kicked out by the thermal noise. The time averages of the uncorrelated and the correlated system are well-explained by the Boltzmann law stated in Eq.~(\ref{eq:BL}). The total simulation used a $T=4e6$  steps to compare the results with the Boltzmann distribution, though only a fraction is shown in the upper panel for clarity.}
\label{fi:figure7}
\end{figure*}
Removing the diagonal entries by making $J_{ii}=0$ and multiplying the matrix entries by 2, to obtain simple integers, $J_{\rm AND}$ evaluates to:
\begin{equation}
\arraycolsep=1.0pt\def\arraystretch{1.0}
J_{\rm AND}=\left(\begin{array}{cccccccc} 0 & -1 & 0 & 0 & 1 & 1 & 1 & 0\\ -1 & 0 & 1 & 1 & 0 & 0 & 0 & 1\\ 0 & 1 & 0 & 0 & 1 & 1 & -1 & 0\\ 0 & 1 & 0 & 0 & 1 & -1 & 1 & 0\\ 1 & 0 & 1 & 1 & 0 & 0 & 0 & -1\\ 1 & 0 & 1 & -1 & 0 & 0 & 0 & 1\\ 1 & 0 & -1 & 1 & 0 & 0 & 0 & 1\\ 0 & 1 & 0 & 0 & -1 & 1 & 1 & 0 \end{array}\right)
\label{eq:ANDs}
\end{equation}
\normalsize
with the notation, [1-5: auxiliary bit and handle bit, 6:``A'', 7:``B'', 8:``C''].  Following a similar procedure, we use the following $14\times 14$ Full Adder matrix, $J_{\rm FA}$:
\footnotesize
\begin{equation}
\arraycolsep=1.0pt\def\arraystretch{1.0}
J_{\rm FA}\hspace{-0pt}= \hspace{-2pt}\left(\hspace{-3pt}\begin{array}{cccccccccccccc} 0 & 0 & 0 & 0 & 0 & 0 & 0 & 4 & -1 & -1 & -1 & -1 & -2 & -1\\ 0 & 0 & 0 & 0 & 0 & 0 & 4 & 0 & -1 & -1 & 2 & -1 & 1 & -1\\ 0 & 0 & 0 & 0 & 0 & 4 & 0 & 0 & -1 & -1 & -1 & 2 & 1 & -1\\ 0 & 0 & 0 & 0 & 4 & 0 & 0 & 0 & -1 & -2 & 1 & 1 & -1 & 1\\ 0 & 0 & 0 & 4 & 0 & 0 & 0 & 0 & -1 & 2 & -1 & -1 & 1 & -1\\ 0 & 0 & 4 & 0 & 0 & 0 & 0 & 0 & -1 & 1 & 1 & -2 & -1 & 1\\ 0 & 4 & 0 & 0 & 0 & 0 & 0 & 0 & -1 & 1 & -2 & 1 & -1 & 1\\ 4 & 0 & 0 & 0 & 0 & 0 & 0 & 0 & -1 & 1 & 1 & 1 & 2 & 1\\ -1 & -1 & -1 & -1 & -1 & -1 & -1 & -1 & 0 & 0 & 0 & 0 & 0 & 0\\ -1 & -1 & -1 & -2 & 2 & 1 & 1 & 1 & 0 & 0 & -1 & -1 & 1 & 2\\ -1 & 2 & -1 & 1 & -1 & 1 & -2 & 1 & 0 & -1 & 0 & -1 & 1 & 2\\ -1 & -1 & 2 & 1 & -1 & -2 & 1 & 1 & 0 & -1 & -1 & 0 & 1 & 2\\ -2 & 1 & 1 & -1 & 1 & -1 & -1 & 2 & 0 & 1 & 1 & 1 & 0 & -2\\ -1 & -1 & -1 & 1 & -1 & 1 & 1 & 1 & 0 & 2 & 2 & 2 & -2 & 0 \end{array}\hspace{-0pt}\right)
 \label{eq:j_fa}
\end{equation}
\normalsize with the notation,  [1$-$9: auxiliary bits and handle bit, 10: ``$C_{\rm in}$'', 11: ``B'', 12: ``A'', 13: ``S'' 14: ``$C_{\rm out}$''].

These are the J-matrices (AND and FA) that are used for all examples in the paper, except for the AND gate described in Section~\ref{sec:hw}. Fig.~\ref{fi:figure10} shows the ``truth table'' operation of the Full Adder where all input/output terminals are ``floating'' using the J-matrix of Eq.~(\ref{eq:j_fa}), showing excellent quantitative agreement with the Boltzmann distribution of Eq.~(\ref{eq:BL}) at steady state even for the undesired peaks of the truth table.  

Note that this prescription for [J] is similar to the principles developed originally for Hopfield networks (\cite{personnaz1986collective}, and Eq.~(4.20) in \cite{amit1992modeling}). However, other approaches are possible along the lines described in the context of Ising Hamiltonians for quantum computers \cite{biamonte2008nonperturbative}. We have tried some of these other designs for [J] and many of them lead to results similar to those presented here. For practical implementations, it will be important to evaluate different approaches in terms of their demands on the dynamic range and accuracy of the weight logic.

\begin{figure*}[t!]
\includegraphics[width=0.75\linewidth]{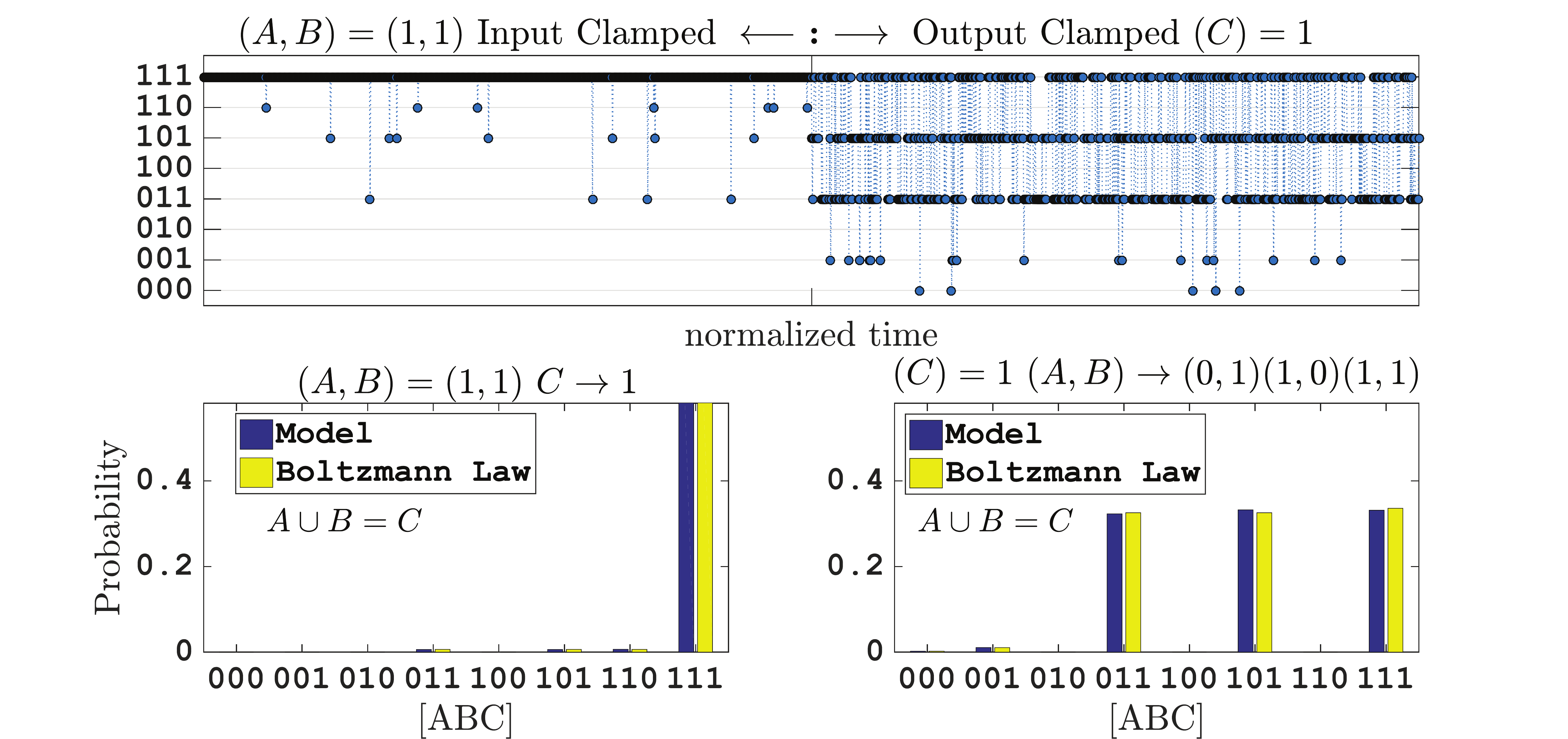}
\caption{\textbf{Implementing a Boolean function and its inverse}: The input or output terminals of an appropriately interconnected network of p-bits can be ``clamped'' to perform a specific logic operation or its \textit{inverse}. In this example, the input bits (A,B) of an OR Gate are clamped to be +1, forcing the output bit C to be 1, during the first phase of operation ($t<t_0$). In the second phase of operation ($t>t_0$), the output of the OR gate C is clamped to the value +1, which is consistent with three different combinations of (A,B). As shown in the time response and the long-time histogram plots, all three possibilities emerge with equal probability, demonstrating the ``inverse'' OR operation. In each case, the expected probabilities from the Boltzmann Law (Eq.~(\ref{eq:BL})) closely match those produced by the generic model, Eq.~(\ref{eq:sigmoid}-\ref{eq:weight}) after running the system for one million steps, only a fraction is shown in the upper panel for clarity.}
\label{fi:figure8}
\end{figure*}

\begin{figure}[t!]
\includegraphics[width=.90\linewidth]{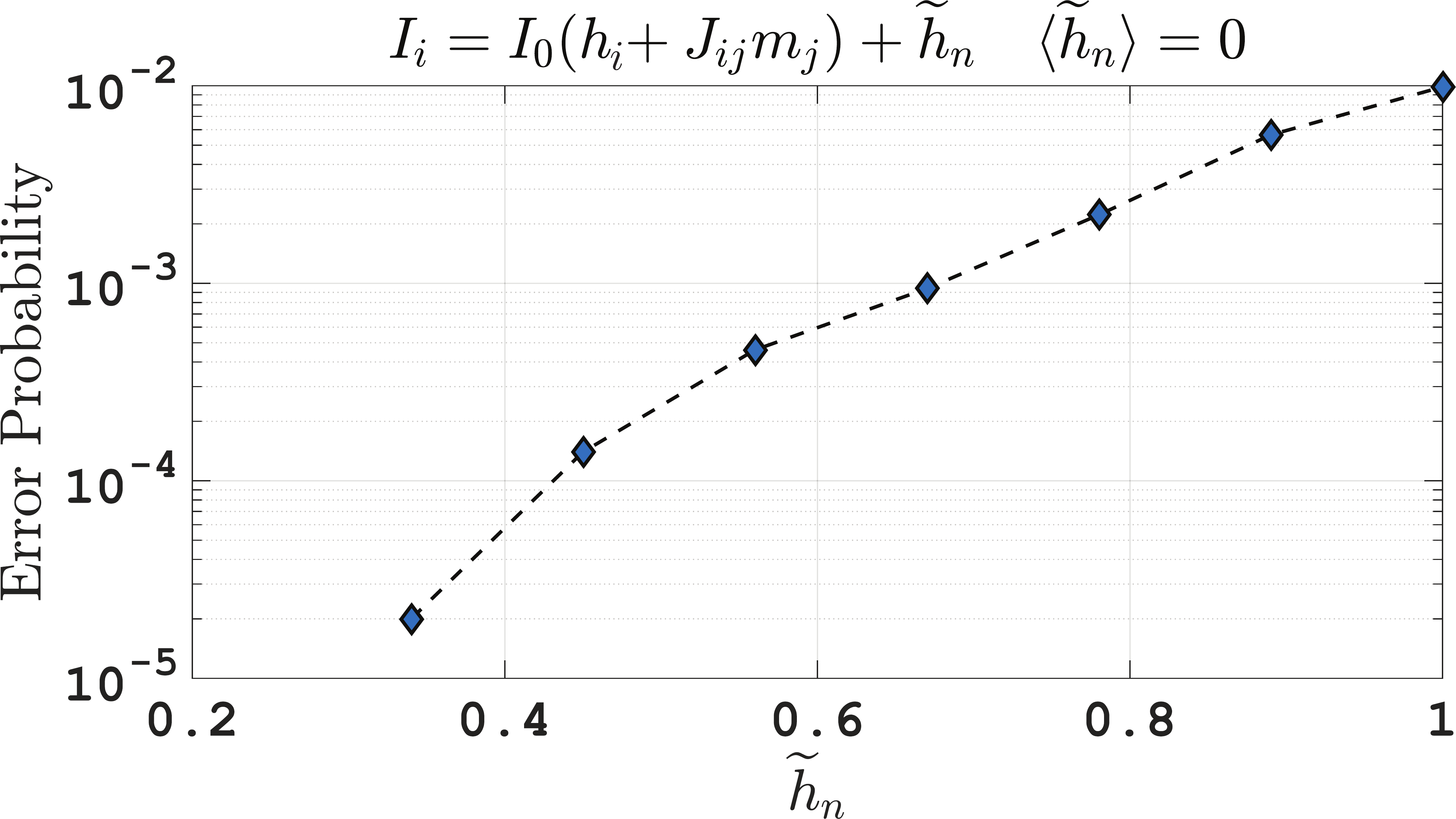}
\caption{\textbf{Noise Tolerance of AND}:  The probability of a wrong output for an (AND) gate (Eq.~\ref{eq:ANDs}) operated with clamped inputs is investigated in the presence of a random noise field which enters Eq.~(\ref{eq:weight}) as indicated in the figure. The noise is assumed to be uniformly distributed over all p-bits in a given network, and centered around zero with magnitude $\pm \tilde{h}_n$, where $(I_0=2, h_i=\pm 1)$.  Each gate is simulated 50000 times for $T$=100 time steps to produce an error probability for a given noise value, and the maximum peak produced by the system is assumed to be an output that can be read with certainty. The system shows robust behavior even in the presence of large levels of noise.}
\label{fi:figure9}
\end{figure}

\textit{Description of universal model:} Once a J-matrix and the h-vector are obtained for a given problem, the system is initialized by randomizing all $m_i$ at time, $t=t_0$. First, the  current (voltage) that  a given p-bit ($m_i$) feels due to the other coupled $m_j$ is obtained from Eq.~(\ref{eq:weight}), and the $m_i$ value is updated according to Eq.~(\ref{eq:sigmoid}). Next the procedure is repeated for the remaining p-bits by finding the current they receive due to all other $m_i$ using the \textit{updated} values of $m_i$. For this reason, the order of updating was chosen randomly in our models and we found that the order of updating has no effect in our results. However, updating the p-bits in \textit{parallel} leads to incorrect results. These two observations are well-known in the context of Hopfield networks  and Boltzmann Machines \cite{aiyer1990theoretical,suzuki2013chaotic,Hinton2007}. This type of serial updating corresponds to the ``asynchronous dynamics'' \cite{hopfield1982neural,amit1992modeling}. We note that the hardware implementation discussed in this paper naturally leads to an asynchronous updating of p-bits in the absence of a global clock signal.  We have set up an online simulator based on this model in Ref.~\cite{brian_sim} so that interested readers can simulate some of the examples discussed in this paper.

Fig.~\ref{fi:figure7} shows the time evolution of an AND based on Eq.~(\ref{eq:ANDs}). Initially for $t<t_0$ the interaction strength is zero ($I_0=0$), making the pseudo-temperature of the system infinite and the network produces uncorrelated noise visiting each state with equal probability. In the second phase ($t>t_0$), the interaction strength is suddenly increased to $I_0=2$, effectively ``quenching'' the network by reducing the temperature. This correlates the system such that only the states corresponding to the  truth table of the AND gate are visited, each with equal probability when a long time average is taken. The average probabilities in each phase  quantitatively match the Boltzmann Law defined by Eq.~(\ref{eq:BL}).

In Fig.~\ref{fi:figure8}, we show how a correlated network producing a given truth table can be used to do directed computation analogous to standard CMOS logic. An OR gate is constructed by using the same [J] matrix for an AND gate, but with a negated handle bit. By ``clamping'' the input bits of an OR gate ($t<t_0$) through their bias terminals, $h_i$, to (A,B)=(+1,+1),  the system is forced to only one of the peaks of the truth table, effectively making C=1.

\begin{figure}[t!]
\includegraphics[width=.95\linewidth]{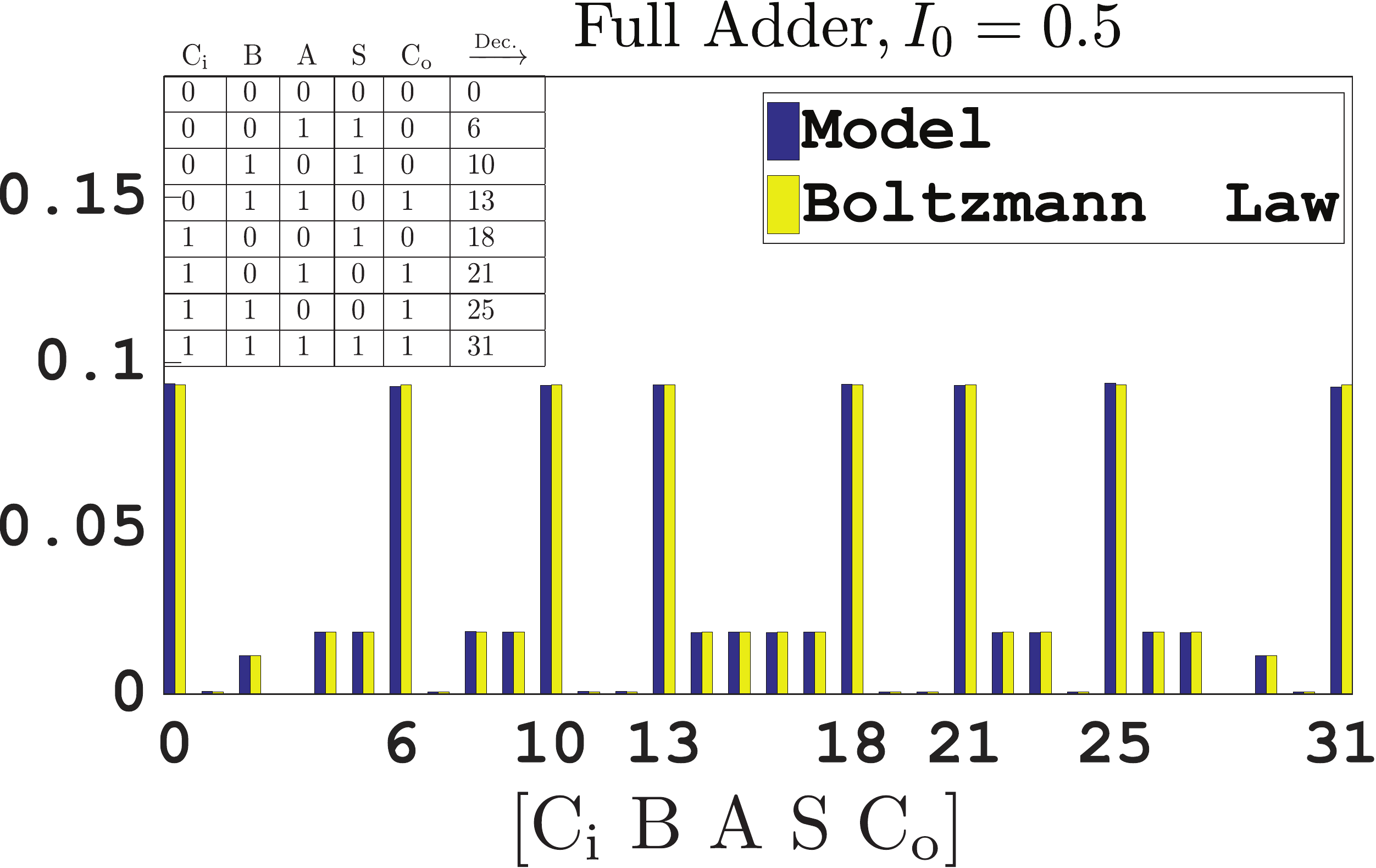}
\caption{\textbf{Full Adder}: Full Adder in the truth table mode, where all inputs and outputs are floating, calculated using $J_{FA}$ from Eq.~(\ref{eq:j_fa}), with $I_0=0.5$. The statistics are collected for $T=10^6$ steps, and each terminal output is then placed in the histogram. The states are numbered using the decimal number corresponding to the binary number $\rm [C_i \ A \ B \ S \ C_o]$. The decimal numbers corresponding to the truth table are shown in the inset, and these match the location of the taller peaks in the histogram. Note that the Boltzmann distribution (Eq.~(\ref{eq:BL})) quantitatively matches the model even for the suppressed peaks.  }
\label{fi:figure10}
\end{figure}
\begin{figure*}[t!]
\includegraphics[width=.99\linewidth]{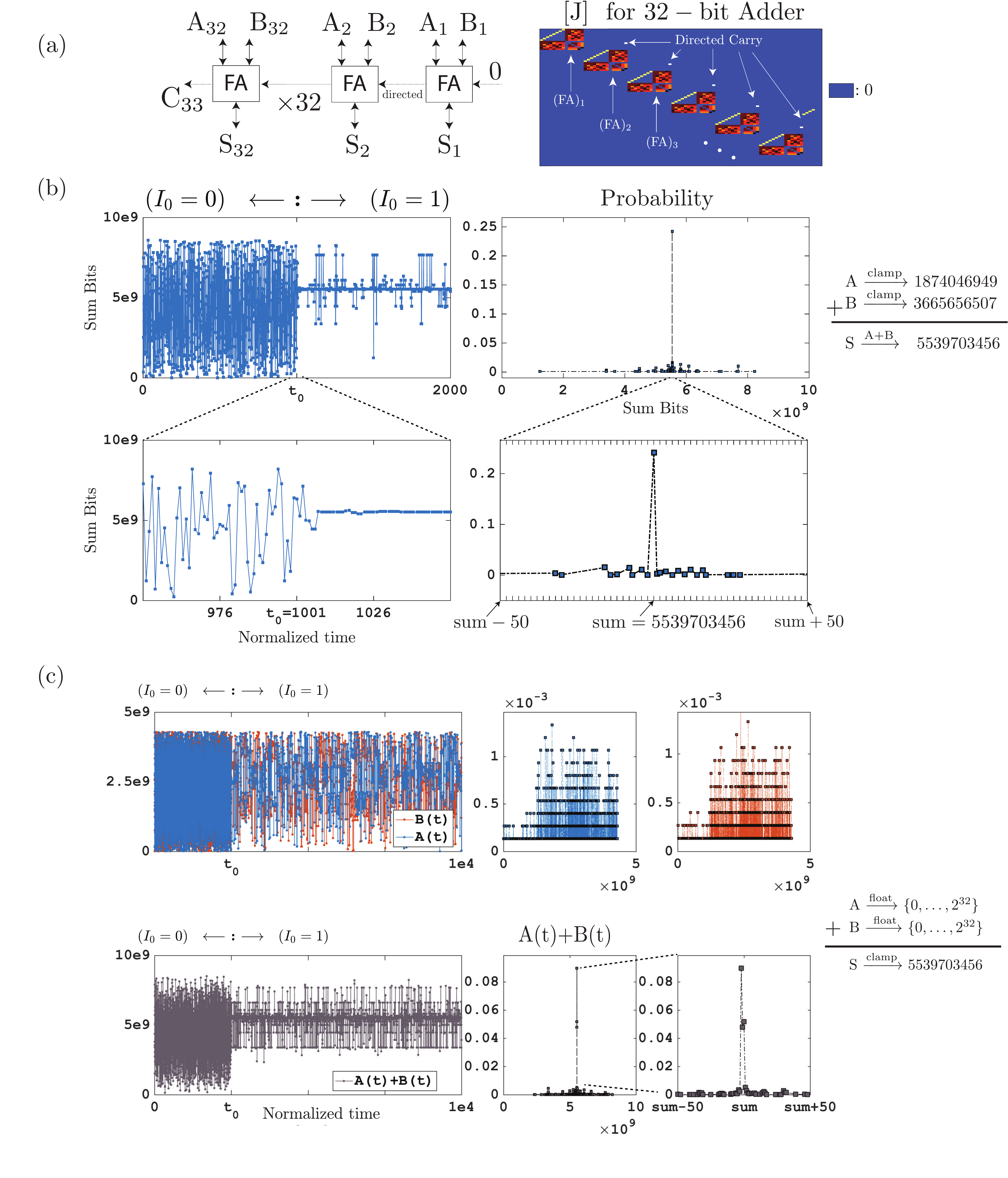}
\caption{\textbf{32-bit Ripple Carry Adder (RCA)}: (a) A 32-bit Ripple Carry Adder (RCA) is designed using individual Full Adder (FA) units with the carry bit designed as a \textit{directed} connection from the least significant bit to the most significant bit. The overall J-matrix for a 32-bit adder J-matrix is shown, and it is quite sparse and quantized. (b) For $t < t_0, I_0 = 0$ and the sum fluctuates randomly. At $t=t_0, I_0$ is suddenly increased, and the adder converges on the correct result for two random inputs A and B. The distribution of 1000 data points ($t>t_0)$ show a single peak with 24$\%$ probability of time spent in the correct state (not including the uncorrelated time points for $t<t_0$). (c) Even though the connections between the Full Adder units are directed, the system performs the inverse function as well. When the output (S) is clamped to a fixed number, the inputs (A) and (B) fluctuate in a correlated manner to make A+B=S when $I_0=1$. Note the broad distributions of A and B (collected for $t>t_0$) as compared to the extremely sharp distribution of A+B.}
\label{fi:figure11}
\end{figure*}
\begin{figure}[t!]
\includegraphics[width=.99\linewidth]{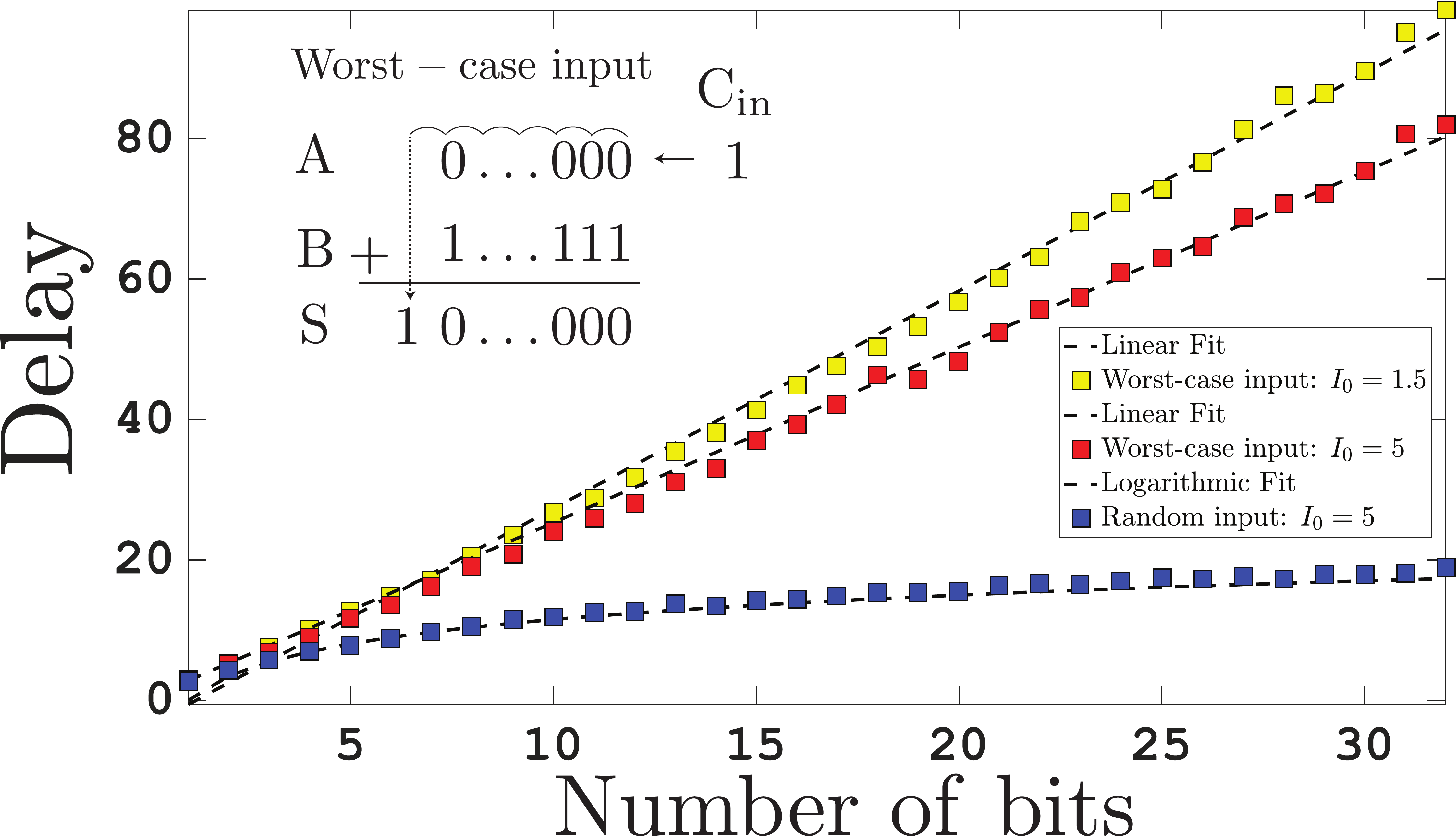}
\caption{\textbf{Ripple Carry Adder delay}: The delay of the RCA  as a function of number of bits in the Ripple Carry Adder (RCA) is shown. The worst case input combination generates a carry that propagates all the way through bit-1 to bit-N, and has a linear dependence on the number of bits, exhibiting O(n) complexity. When the inputs are random, the delay increases logarithmically.  The delay is defined to be the time it takes for the network to reach the mode of the array for $T$=200 after getting quenched at t=0. Each point is an average of 500 trials with random initial conditions for an $I_0=1.5$, and the mode of the array was exactly equal to the arithmetic sum of the inputs in each case. The worst-case inputs are A=$0 \ldots 000$ and  B=$ 1 \ldots 111$ with an input carry ($\rm C_{in})$ of 1. Results show a weak $I_0$ dependence.}
\label{fi:figure12}
\end{figure}

The PSL gates however exhibit a remarkable difference with standard logic gates, in that inputs and outputs are on an equal footing. Not only do clamped inputs give  the corresponding output, \textit{a clamped output gives  the corresponding input(s).} In the second phase ($t>t_0$) the output of the OR gate is clamped to +1, that produces three possible peaks for the input terminals, corresponding to various possible input combinations that are consistent with the clamped output (A,B)=(0,1),(1,0) and (1,1). The probabilistic nature of PSL allows it to obtain multiple solutions (Fig.~\ref{fi:figure8}c). It also seems to make the results more resilient to \textit{unwanted} noise due to stray fields that are inevitable in physical implementations as shown in Fig.~\ref{fi:figure9}. Here, we simulate an AND gate in the presence of a normally distributed random noise that enters the bias fields of each p-bit and define the computation to be faulty, if the mode (most frequent value) of the output bit is not consistent with the programmed input combinations after $T = 100$ time steps. We observe that even large levels of uncontrolled noise produces correct results with high probabilities. 

Fig.~\ref{fi:figure10} shows the design of a Full Adder (FA) with the 8-line truth table shown. There are three inputs in all, two from the numbers to be added, and one carry bit from previous FA. It produces two outputs, one the sum bit and the other a carry bit to be passed on to the next FA. The probabilities of different states are  calculated using $J_{FA}$ from Eq.~(\ref{eq:j_fa}), with $I_0=0.5$ in the truth table mode, where all inputs and outputs are floating and the states are numbered using the decimal number corresponding to the binary word $\rm [C_i \ A \ B \ S \ C_o]$. The decimal numbers corresponding to the truth table are shown in the inset, and these match the location of the taller peaks in the histogram. Note that the Boltzmann distribution (Eq.~(\ref{eq:BL})) quantitatively matches the model even for the suppressed peaks. A higher $I_0$ reduces these suppressed peaks  further. The statistics are collected for $T=10^6$ steps, and each terminal output is then placed in the histogram.
\begin{figure*}[t!]
\includegraphics[width=0.9\linewidth]{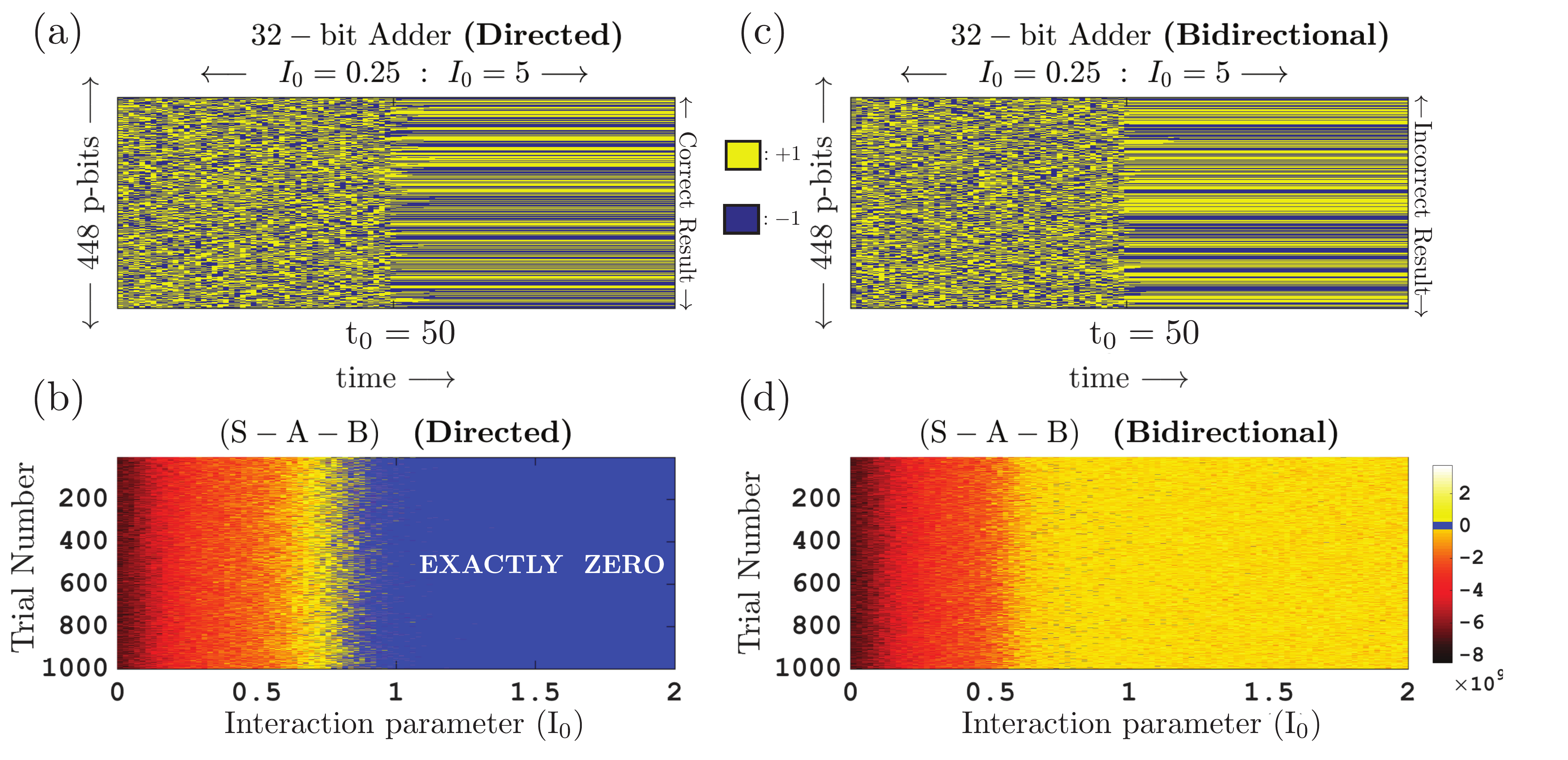}
\caption{\textbf{Accuracy of 32-bit adder, directed versus bidirectional:} The results are shown for the adder operating in a subtractor mode, clamping one (random) 32-bit input (A) and a (random) 33-bit output ($C_{\rm out}$+ S), and observing the other 32-bit input B which should provide the difference S$-$A. (a): Colormap of the binary state of each of the 448 p-bits comprising the directed adder as a function of time with the interaction parameter $I_0$ suddenly increased from 0.25 to 5 at $\rm t_0$=50. For low values of $I_0$ at t$<$50, the collection of p-bits is like a molten liquid which is quenched at $\rm t_0=50$ into a solid. (b) Surprisingly this solid corresponds to a ``perfect crystal'' in each of the 1000 trial experiments, with S$-$A$-$B exactly equal to zero (Dark blue). (c) Same as (a) but for a bidirectional adder. Here too the ``liquid'' quenches to a solid at $\rm t_0=50$, but in this case the resulting ``solid'' is full of defects (with hardly any zeros), with S$-$A$-$B $\neq 0$, yielding a different wrong result for each trial as evident from (d).  For (c) and (d) The colorbar is modified to have a dark blue color corresponding to exactly zero. S,A,B are taken to be the statistical mode of the 100$\times$1 array obtained at the end of each trial.}
\label{fi:figure13}
\end{figure*}
\section{Directed Networks of Boltzmann Machines}
\label{sec:directed}

When constructing larger circuits composed of individual Boltzmann machines, the reciprocal nature of the Boltzmann machine often interferes with the directed nature of computation that is desired. It seems advisable to use a hybrid approach. For example in constructing a 32-bit adder we use Full-Adders (FA) that are individually BMs with symmetric connections, $J_{ij} = J_{ji}$. But when connecting the carry bit from one FA to the next, the coupling element $J_{ij}$ is non-zero in only one direction from the least significant to the most significant bit. This directed coupling between the components distinguishes PSL from purely reciprocal Boltzmann machines. Indeed, even the Full Adder could be implemented not as a Boltzmann machine but as a directed network of more basic gates. But then it would lose its invertibility. On the other hand, the directed connection of BM Full Adders largely preserves the invertibility of the overall system as we will show.
  \vspace{0em}
\subsection{32-bit Adder/Subtractor}
  \vspace{0em}
Fig.~\ref{fi:figure11} shows the operation of a 32-bit adder that sums two 32-bit numbers A and B to calculate the 33-bit sum S. In the initial phase ($t < t_0$) we  have $I_0=0$ corresponding to infinite temperature so that the sum bits (S) fluctuate among $2^{33} \approx$ 8 billion possibilities. With $I_0$ = 1, Fig.~\ref{fi:figure11} shows that the correct answer has a probability of $\approx 12 \%$ which is much lower than the $\approx 100\%$ that can be achieved with larger $I_0$ values (as in Fig.\ref{fi:figure13} a-c with $I_0$=5). Nevertheless  the peak is unmistakable as evident from the expanded scale histogram and the correct answer is extracted from the majority vote of $T$=100 samples as shown in Fig.~\ref{fi:figure13}. This ability to extract the correct answer despite large fluctuations is a general property of probabilistic algorithms.


Interestingly, although the overall system includes several unidirectional connections, it seems to be able to perform the inverse function as well. With A and B clamped it calculates S=A+B as noted above. Conversely with S clamped, the input bits A and B fluctuate in a correlated manner so as to make their sum sharply peaked around S. Fig.~\ref{fi:figure11} shows the  time evolution of the input bits that have broad distributions spanning a wide range. Initially, when $I_0$ is small, the sum of A and B  also shows a broad distribution, but once $I_0$ is turned up to 1, the distributions of  A and B get strongly correlated making the distribution of A+B sharply peaked around the fixed value of S. It must be noted that the 32-bit adder shown in Fig.~\ref{fi:figure11} is not like standard digital circuits which are not invertible.  The demonstration of such an invertible 32-bit adder could be practically significant, since binary addition is noted to be the most fundamental and frequently used operation in digital computing \cite{liu2003algorithmic}.

\textit{Delay of Ripple Carry Adder}: Just as in CMOS-based Ripple Carry Adders, the delay of the p-bit based RCA is a function of the inputs A and B. In Fig.~\ref{fi:figure12} we have systematically studied the worst-case delay of the p-bit based Ripple Carry Adder (RCA) as a function of increasing bit size. We selected a ``worst-case'' combination that results in a carry that needs to be propagated from bit 1 to bit N which results in a linear increase in the delay, exhibiting O(n) complexity with input size similar to CMOS implementations \cite{uma2012area}. When the inputs are random, the delay seems to increase sub-linearly. The system is quenched at t=0 for different interaction parameters $I_0$ and the delay is defined to be the time it takes for the system to settle to the mode of the array for  $T$=200. An error check has been carried out separately to ensure the calculated sum (mode) is always exactly equal to the expected sum. For random inputs the 32-bit adder is close to 20 time steps, in accordance with the example shown in Fig.~{\ref{fi:figure11}}.  

\textit{Digital accuracy AND logical invertibility:} The striking combination of accuracy and invertibility is made possible by our hybrid design, whereby the individual Full Adders are Boltzmann Machines, even though their connection is directed. Our 32-bit adder is more like a  \textit{collection of interacting particles} than like a digital circuit as evident from Fig.~\ref{fi:figure13}a which shows a colormap of the binary state of each of the 448 p-bits as a function of time with the interaction parameter $I_0$ suddenly increased from 0.25 to 5 at $\rm t_0=50$, thereby quenching a ``molten liquid'' into a ``solid''. Nevertheless it shows the striking accuracy of a digital circuit, with S$-$A$-$B exactly equal to zero in each of the 1000 trials as shown in  Fig.~\ref{fi:figure13}b. We do not expect a ``molten liquid'' to be quenched into a ``perfect crystal'' every time. Instead, we would expect a ``solid full of defects'' with different non-zero values for S$-$A$-$B in each trial. That is exactly what we get if the carry bits are bidirectional as in a fully BM implementation (Fig.~\ref{fi:figure13}d). 

\begin{figure*}[t!]
{\includegraphics[width=0.90\linewidth]{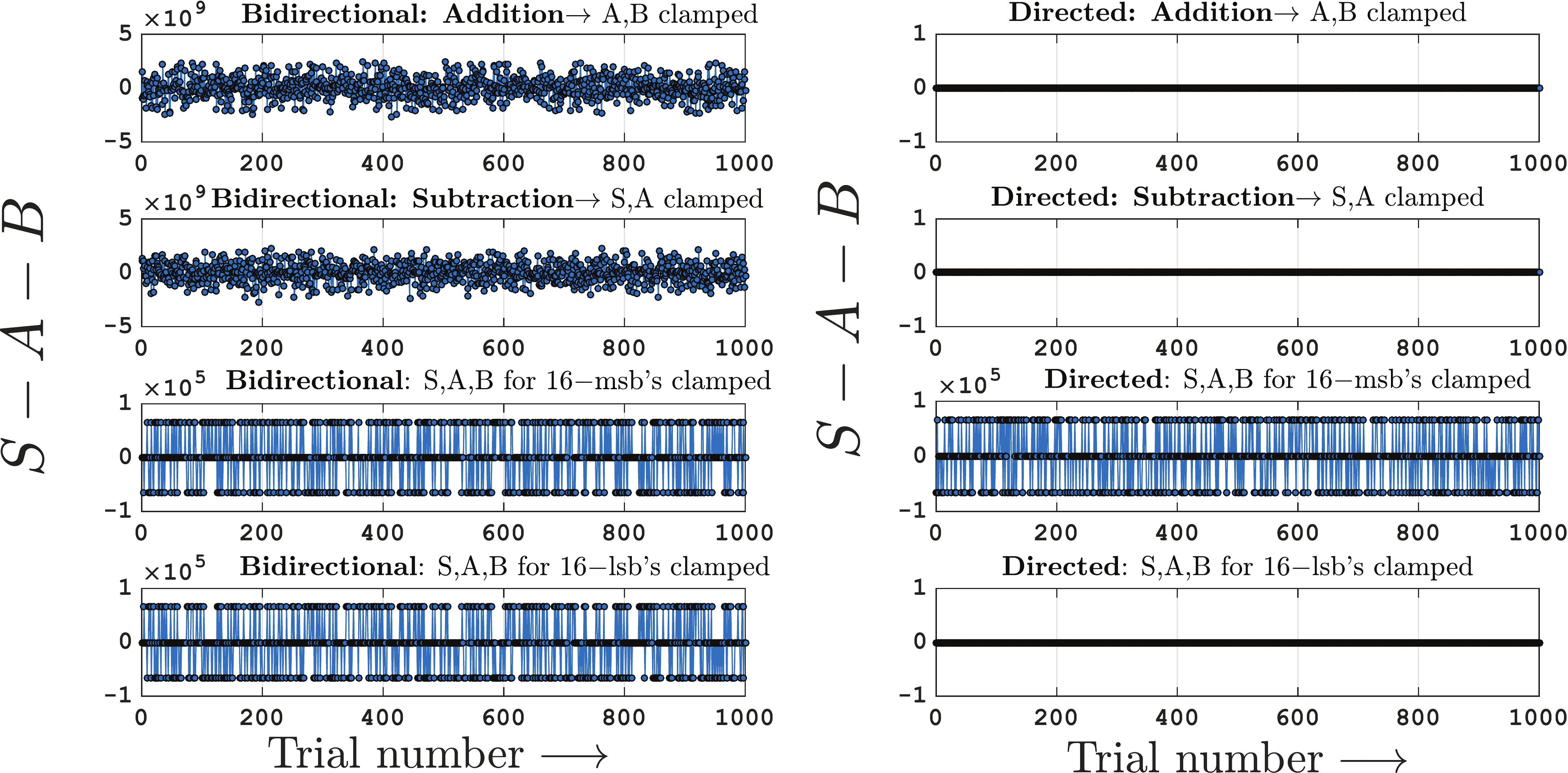}}
\caption{\textbf{Invertibility of 32-bit adder, directed vs bidirectional:} An adder that provides the sum S of two 32-bit numbers A and B: $S=A+B$. The left panel shows the adder implemented with bidirectional carry bits, while the right panel shows one with carry bits directed from the least significant to the most significant bit.  Four different modes are shown with (i) A and B clamped (Addition), (ii) S and A clamped (Subtraction), (iii) A, B and S for the 16 most significant bits (msb) clamped, and (iv) A, B and S for the 16 least significant bits (lsb) clamped. Note that that bidirectional implementation shows very large errors for all modes of operation. The directed implementation works perfectly for both the adder and the subtractor modes. It also works if we clamp the least significant bits, but not if we clamp the most significant bits. Correlation parameter $I_0=1$, $T =100$ steps for all trials. S,A,B are taken to be the mode (most frequent value) of the 100$\times$1 array obtained at the end of each trial. Clamped inputs are random 32-bit words for each trial, for a total of 1000 trials.}
\label{fi:figure14}
\end{figure*}
\begin{figure}
{\includegraphics[width=0.99\linewidth]{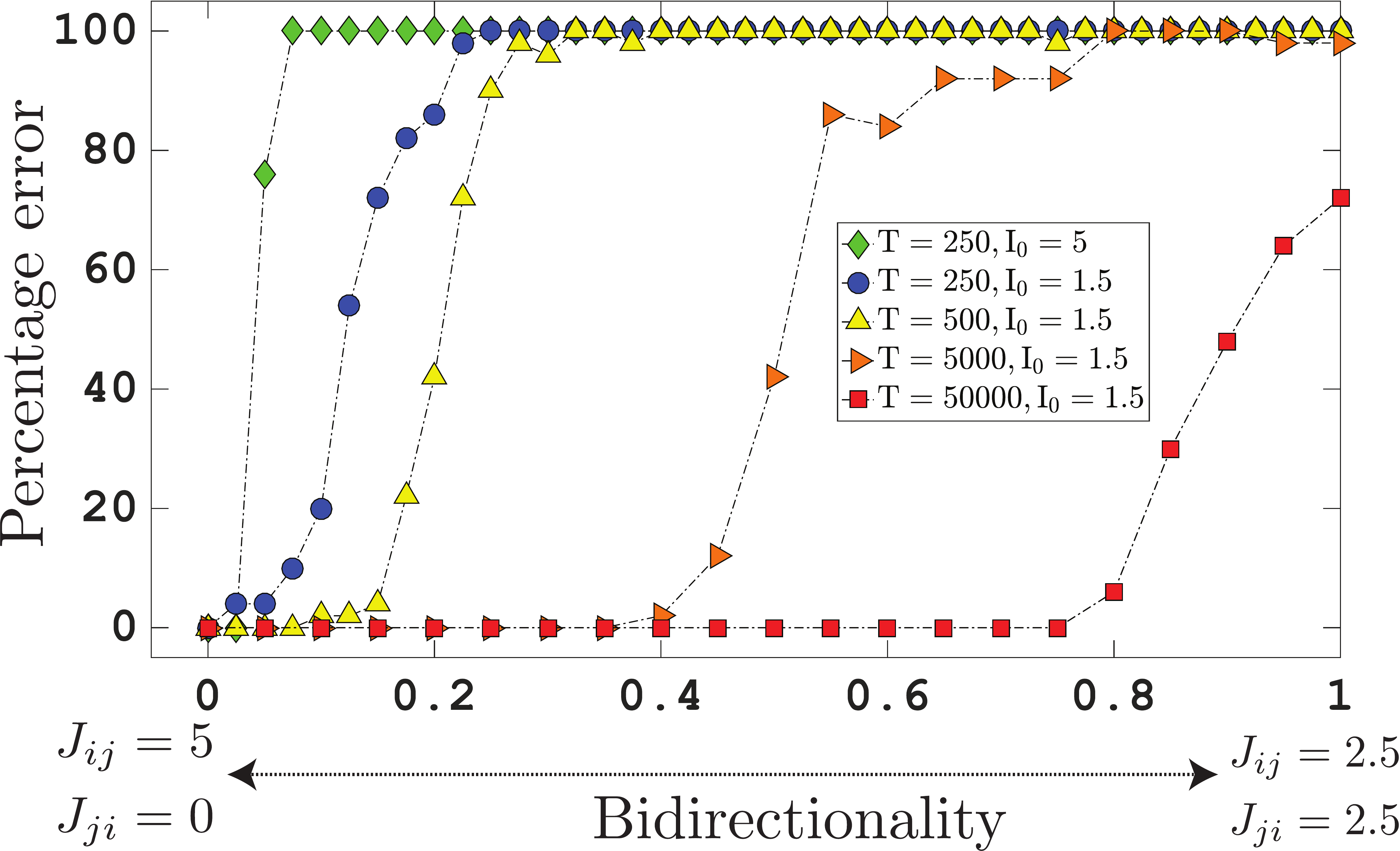}}
\caption{\textbf{Error versus bidirectionality:} The degree of bidirectionality $J_{ji}/J_{ij}$ of the carry-out (j) to carry-in (i) link between the Full Adders is systematically varied while keeping the sum $J_{ij}+J_{ji}$ constant. In each case the sum is obtained from the statistical mode (or majority vote) of $T$ time samples over 50 trials. The y-axis shows the fraction of trials that yield the wrong result. Note that for large $I_0$ and small $T$, error-free operation is obtained only if bidirectionality is close to zero similar to standard digital circuits. But with $I_0$ = 1.5 and $T$=50,000, error-free operation (at least for 50 trials) is obtained even with $\approx 75 \%$ bidirectionality.}
\label{fi:figure15}
\end{figure}

Note however, that this digital accuracy is achieved while maintaining the property of invertibility that is absent in digital circuits. Fig.~\ref{fi:figure13} is not for direct mode operation, but for the adder operating in reverse mode as a subtractor. It might be expected that the directed connection of carry bits from the less significant to the more significant bit could lead to a loss of invertibility. To investigate this point, we show the error S$-$A$-$B as a function of trial number (Fig.~\ref{fi:figure14}) for four different modes of operation with (i) A and B clamped (Addition), (ii) S and A clamped (Subtraction), (iii) A, B and S for the 16 most significant bits (msb) clamped, and (iv) A, B and S for the 16 least significant bits (lsb) clamped. The fully bidirectional implementation shows very large errors for all modes of operation. The directed implementation, on the other hand, works perfectly for both the adder and the subtractor modes. It also works if we clamp the least significant bits, but not if we clamp the most significant bits. This seems reasonable since we expect to be able to control a flow by making changes upstream (lsb), but not downstream (msb).

\textit{Partial directivity:} So far in our examples we have only considered fully directed ($J_{ij}=2 \ J_0, J_{ji}=0$) or fully bidirectional ($J_{ij}=J_0, J_{ji}=J_0$)  carry bits when connecting the individual Full Adders. In Fig.~\ref{fi:figure15} we systematically analyze the effects of partial directivity in the operation of a 32-bit adder. We observe that the 32-bit adder operates correctly  even when there is  large degree of bidirectionality  ($J_{ji}=J_{ij} \times 0.75$) provided that the system is allowed to run for a long time, $T=50000$, in stark contrast with the fully directed case that could resolve the right answer within $T=100$, shown in Fig.~\ref{fi:figure14}b. Decreasing the time steps systematically increases the error. Increasing the correlation parameter while keeping $T$ constant also seems to adversely affect the bidirectional designs, that might be getting the system stuck in local minima. 
\begin{figure*}[t!]
\includegraphics[width=.99\linewidth]{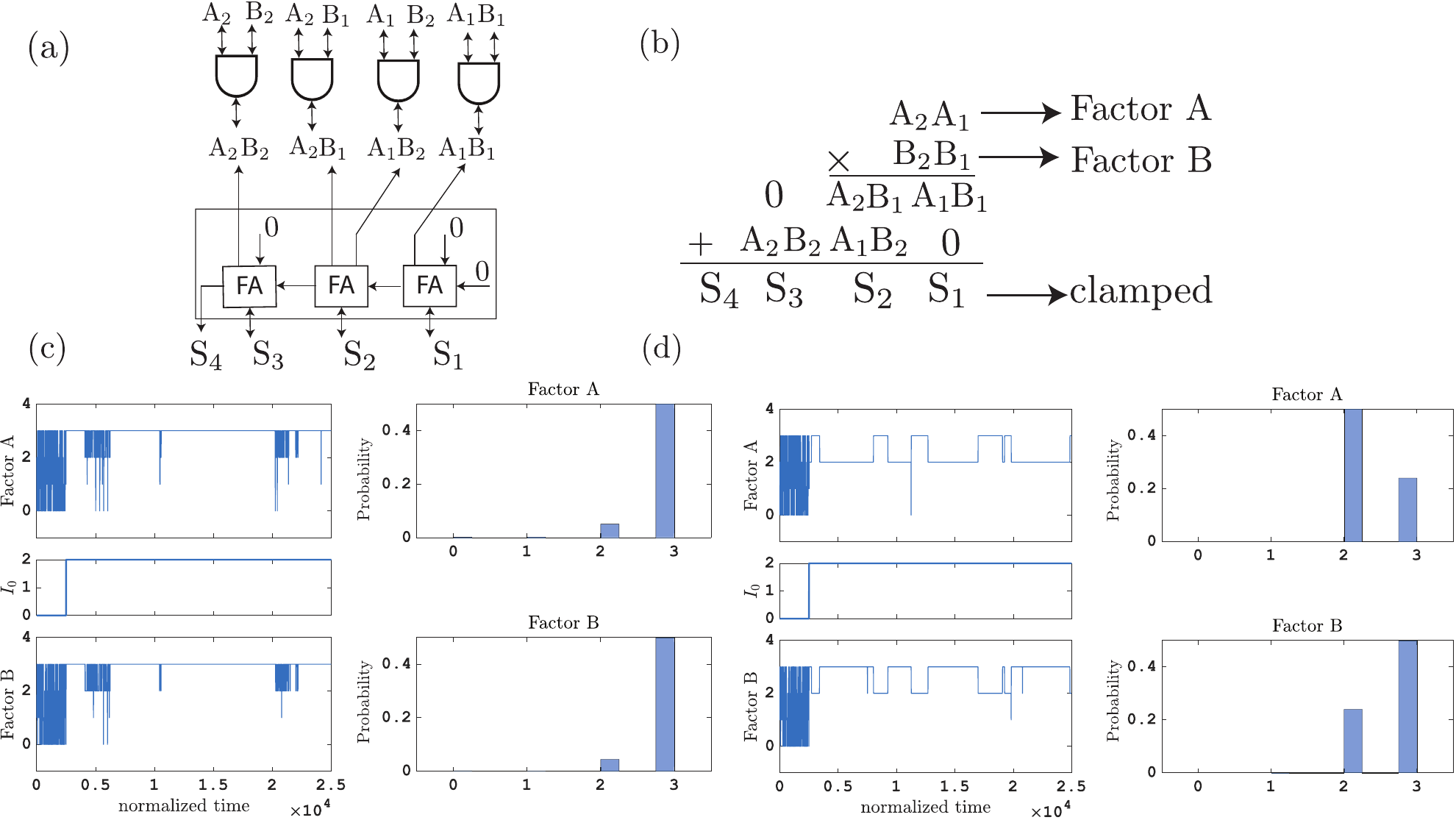}
\caption{\textbf{Factorization through inverse multiplication}: The reversibility of PSL allows the operation of integer factorization using a binary multiplication circuit implemented using the principles of digital logic using AND gates and Full Adders (FA) as shown in (a). The output nodes of a 4-bit multiplier are clamped to a given integer, and the system produces the only consistent factors of the product at the input terminals, probabilistically. The interaction parameter $I_0$  is suddenly increased to a saturation value of  2, and held constant as shown. (b) The output terminal is clamped to 9 and is factored into $3\times 3$, note that $9 \times 1 $ is not an achievable  solution in this setup since encoding 9 requires 4-bit inputs in binary, whereas inputs  are limited to 2-bits. (c) The output terminal is clamped to 6 and after being correlated, the factors  cross-oscillate between 2 and 3. In both cases the histogram is obtained by counting outputs after $t>t_{\rm total}/2=1.25\times 10^4$ time steps to collect statistics after the system is thermalized.}
\label{fi:figure16}
\end{figure*}

\textit{Directionality and computation time, 2 p-bit model}: The qualitative relation between $I_0$, $T$ and bidirectionality $J_{12}/J_{21}$ described above is derived from extensive numerical simulations based on Eq.~\ref{eq:sigmoid}-\ref{eq:weight}. However, the  broad features can be understood from a model involving just two p-bits, 1 and 2, with
\[h=
\begin{bmatrix}
0\\
0
\end{bmatrix}
\ \  \mathrm{and}  \ \  J=
\begin{bmatrix}
0 & J_{12} \\
J_{21} & 0
\end{bmatrix}\]
\noindent It is straightforward to write a master equation describing the time evolution of the probabilities of different configurations:
\[\frac{d}{dt}
\begin{bmatrix}
P_{11}\\
P_{10}\\
P_{01}\\
P_{00}
\end{bmatrix}
=[W]
\begin{bmatrix}
P_{11}\\
P_{10}\\
P_{01}\\
P_{00}
\end{bmatrix}\]
\noindent $W$ being the transition matrix \cite{amit1992modeling}, $P_{00}$ representing the probability of both p-bits being $-1$, $P_{11}$ both being $+1$, and so on. We can write two matrices $W_1$ and $W_2$ describing the updating of p-bits 1 and 2 respectively:
\[ 
W_1= {\bbordermatrix{(1,2) ~& \rm (11) & \rm (10) & \rm (01) & \rm (00) \cr 
\rm (11)   & p & 0 & p & 0\cr 
\rm (10)   & 0 & {\overline{p}} & 0 & {\overline{p}} \cr 
\rm (01)   & {\overline{p}} & 0 & {\overline{p}}  & 0 \cr
\rm (00)   & 0 & p & 0  & p }}\]
\vspace{-10pt}
\[
W_2= {\bbordermatrix{(1,2) ~& \rm (11) & \rm (10) & \rm (01) & \rm (00) \cr 
\rm (11)   & q & q & 0 & 0\cr 
\rm (10)   & {\overline{q}} & {\overline{q}} & 0 & 0 \cr 
\rm (01)   & 0 & 0 & {\overline{q}}  & {\overline{q}} \cr
\rm (00)   & 0 & 0 & q  & q }}
\]
\noindent where $W(i,j)$ represents the probability that state $(j)$ makes a transition to state $(i)$, and $\bar{p}=1-p$, $\bar{q}=1-q$. $p$ and $q$ are  obtained from Eq.~\ref{eq:sigmoid}-\ref{eq:weight}:
\[ p=\frac{1}{2}(1+\mathrm{tanh}(I_0 (J_{12}+h_{1})))=\frac{1}{2}(1+\mathrm{tanh}(I_0 J_{12})) \]\vspace{-20pt}
\[ q=\frac{1}{2}(1+ \mathrm{tanh}(I_0 (J_{21}+h_{2})))=\frac{1}{2}(1+\mathrm{tanh}(I_0 J_{21}))\]

The overall transition matrix $W$ is given by $W_2 \times W_1$ or $W_1\times W_2$ depending on which bit is updated first. Either way the matrix $W$ has four eigenvalues $\lambda_1= 1, \lambda_2= 0, \lambda_3=0$ and $ \lambda_4= (2p-1)(2q-1) =  \mathrm{tanh}(I_0 J_{12}) \times \mathrm{tanh}(I_0 J_{21})$ and the corresponding eigenvectors evolve with time $\sim \lambda^T$. 

The components corresponding to $\lambda$=0 decay instantaneously while the eigenvector corresponding to $\lambda$=1 is the stationary result representing the correct solution. But for the system to reach this state, we have to wait for the fourth eigenvector corresponding to $\lambda_4$ to decay sufficiently. A fully directed network has $J_{21}$ =0, so that $\lambda_4=0$ and the system quickly reaches the correct solution. But in a bidirectional network with $J_{12}=J_{21}$, the fourth eigenvalue can be quite close to one, especially for large $I_0$ and take an exponentially long time to decay, as $\lambda^T = \exp( T \  \mathrm{ln} \ \lambda )\approx \exp(-T (1-\lambda))$ when  $\lambda$ is close to 1.

This 2 p-bit model provides some insight into our general observation that directivity can be used to obtain accurate answers quickly. However, depending on the problem at hand it may be desirable to retain some degree of bidirectionality, since full directivity does lead to some loss of invertibility as seen for one set of inputs in Fig.~\ref{fi:figure14}. An example of a partially directed p-bit network is discussed in the next section.

\subsection{4-Bit Multiplier / Factorizer}

 Fig.~\ref{fi:figure16} shows how the invertibility of PSL logic blocks can be used to perform integer factorization using a multiplier in reverse. Normally, the factorization problem requires specific algorithms \cite{knuth1976analysis} to be performed in CMOS-like hardware, here we simply use a digital 4-bit multiplier working in \textit{reverse} to achieve this operation. 

Specifically with the output of the multiplier clamped to a given integer from 0 to 15, the input bits float to the correct factors. The interconnection strength $I_0$ is increased suddenly from 0 to 2 at $t=t_0$ (Fig.~\ref{fi:figure16}) and the input bits get locked to one of the possible solutions. For example, when the output is set to 9, both inputs float to 3. With the output set to 6, both inputs fluctuate between two values, 2 and 3. Note that factors like $9=9 \times 1$ do not show up, since encoding 9 in binary requires 4-bits (1001) and the input terminals only have 2-bits. We have checked other cases where factorizing 3 shows both $3\times 1$ and $1\times 3$, and factorizing zero shows all possible peaks since there are many solutions such that $0=0\times 1,2,3$ and so on.

We also kept the same directed connections between the Full Adders for the carry bits, making them a directed network of Boltzmann Machines, similar to the 32-bit Adder. Moreover, we kept a directed connection \textit{from} the Full Adders \textit{to} the AND gates  as shown in Fig.~\ref{fi:figure16}a since the information needs to flow from the output to the input in the case of factorization. The input bits that go to multiple AND gates are ``tied'' to each other with a positive exchange ($J>0$) value much like 2-spins interacting ferromagnetically, however in PSL we envision these interactions to be controlled purely electrically. In this example, we have observed that the system is sensitive to the relative strengths of couplings within the AND gates and between the AND gates and the Full Adders which can also depend on a chosen annealing profile.

The design of factorizers of practical relevance is beyond the scope of this paper. Our main purpose has been to establish how the key  feature of invertibility of p-bits can be creatively used for different circuits with unique functionalities.  The demonstration of  4-bit factorization through reverse multiplication is similar to memcomputing \cite{traversa2017} based on deterministic memristors.  Note, however, that the building blocks and operating principles of stochastic p-bits and memcomputing \cite{di2016topological} are very different and the only similarity noted here is the fact that both approaches treat  the input and output terminals on an equal footing.

\section{Summary}
 It is generally believed that (1) probabilistic algorithms can tackle specific problems much more efficiently than classical algorithms \cite{ekert1996quantum}, and that (2) probabilistic algorithms can run far more efficiently on a probabilistic computer than on a deterministic computer \cite{feynman1982simulating,ekert1996quantum}. As such, it seems reasonable to expect that probabilistic computers based on robust room temperature p-bits  could  provide a practically useful solution to many challenging problems by rapidly sampling the phase space in hardware.

In this paper we have presented a framework for using probabilistic units or ``p-bits'' as a building block for a probabilistic spin logic (PSL) which is used to implement precise Boolean logic with an accuracy comparable to standard digital circuits, while exhibiting the unique property of invertibility that is unknown in deterministic circuits. Specifically we have:
\begin{itemize}
\item presented an implementation based on stochastic nanomagnets to illustrate the importance of three-terminal building blocks in the construction of large scale correlated networks of p-bits. We emphasize that this is just one possible implementation that is by no means the only one \textbf{(Section~\ref{sec:hw})}.  \vspace{0.0em}

\item presented  an algorithm for implementing Boolean gates as BM with relatively sparse and quantized J-matrix elements, benchmarked their operation against the Boltzmann law, and established their capability to perform not just direct functions but also their inverse  \textbf{(Section~\ref{sec:truth})}, and   \vspace{-0.0em}

\item presented a 32-bit adder implemented as a hybrid BM that achieves digital accuracy over a broad combination of the interaction parameter $I_0$, directionality and the number of samples $T$. This striking accuracy is reminiscent of digital circuits, but it is achieved  while preserving a certain degree of invertibility which is  absent in digital circuits. The accuracy is particularly surprising with high degrees of bidirectionality ($J_{12}= 0.75 \times  J_{21}$) where the system is picking out the one correct answer out of nearly $2^{33} \approx$  8 billion possibilities. This may require a larger number of time samples, but these could be collected rapidly at GHz rates.  \textbf{(Section~\ref{sec:directed})}.
\end{itemize}
We hope these findings will help emphasize a new direction for the field of spintronic and nanomagnetic logic by shifting the focus from stable high barrier magnets to stochastic, low barrier magnets, while inspiring a search for other possible physical implementations of p-bits.
\begin{acknowledgments}
It is a  pleasure to acknowledge many helpful discussions with Behtash Behin-Aein (Globalfoundries) and Ernesto E. Marinero (Purdue University). We thank Jaijeet Roychowdhury (UC Berkeley) for suggesting the phrase ``invertible''. This work was supported in part by C-SPIN, one of six centers of STARnet, a Semiconductor Research Corporation program, sponsored by MARCO and DARPA, in part by the Nanoelectronics Research Initiative through the Institute for Nanoelectronics Discovery and Exploration (INDEX) Center,  and in part by the National Science Foundation through the NCN-NEEDS program, contract 1227020-EEC.
\end{acknowledgments} 
\end{document}